\documentclass[prb,twocolumn,showpacs]{revtex4}

\usepackage{mathptmx}
\usepackage{amsmath,latexsym}
\usepackage{graphicx}
\usepackage[colorlinks,citecolor=blue]{hyperref}
\providecommand{\texorpdfstring}[2]{#1}

\renewcommand*{\k}{\vec{k}}
\newcommand*{\kk}{\vec{k'}}
\newcommand*{\ks}{\vec{k}\sigma}
\newcommand*{\e}{\varepsilon}
\newcommand*{\ek}{\varepsilon_{\vec{k}}}
\newcommand*{\w}{\omega}

\newcommand*{\abs}[1]{\lvert #1\rvert}
\newcommand*{\expect}[1]{\langle #1\rangle}
\newcommand*{\correlation}[2]{\langle\!\langle #1| #2 \rangle\!\rangle}
\newcommand*{\etal}{\emph{et al.}}

\DeclareMathOperator{\re}{Re}
\DeclareMathOperator{\im}{Im}

\begin{document}

\title{Transport properties of heavy-fermion systems}

\author{Claas Grenzebach}
\author{Frithjof B.~Anders}
\author{Gerd Czycholl}
\affiliation{Institut f\"ur Theoretische Physik, Universit\"at Bremen,
             P.O. Box 330 440, D-28334 Bremen, Germany}

\author{Thomas Pruschke}
\affiliation{Institut f\"ur Theoretische Physik,
             Universit\"at G\"ottingen, D-37077 G\"ottingen, Germany}

\date{2 November 2006}

\begin{abstract}
We calculate the temperature dependence of the transport properties of
heavy-fermion systems such as resistivity, optical conductivity,
thermoelectric power, the electronic part of the thermal conductivity,
and the ``figure of merit.'' The one-particle properties of the
periodic Anderson model are obtained within dynamical mean-field theory 
for the  paramagnetic phase using Wilson's numerical renormalization
group and the modified perturbation theory as impurity solvers. We discuss
the dependence of the transport properties on the band filling, valence,
and Coulomb correlation $U$. The typical experimental findings can be
reproduced and understood, in particular the temperature dependence of
the resistance and the thermoelectric power and their absolute magnitude
for both metallic heavy-fermion systems and Kondo insulators. For large
values of $U$, we find a negative Seebeck coefficient $S(T)$ for an
intermediate-temperature regime as observed in $S(T)$ of CeCu$_2$Si$_2$.
We analyze different estimates for possible characteristic low-temperature
scales of the lattice. Our results indicate a one-parameter scaling of
thermodynamic and some transport properties with a strongly
occupancy-dependent scaling function. This is consistent with a
strong-coupling local Fermi-liquid fixed point of the effective site
governing all low-lying excitations for $T\to 0$ in the paramagnetic phase.
\end{abstract}

\maketitle

\section{Introduction}

For nearly 30 years the investigation of heavy-fermion systems (HFSs) has
been one of the most fascinating and interesting fields in condensed matter
physics both experimentally and theoretically.\cite{Grewe91,Stewart01}
The heavy-fermion phenomenon exists in a number of lanthanide and
actinide compounds and manifests itself in the apparent existence of
quasiparticles with very large effective mass $m^*$ at low
temperatures $T$. This is seen already in the specific heat, the
electronic part of which shows the usual linear temperature dependence
$c = \gamma T$ at very low $T$ for many metallic heavy-fermion systems
but with a $\gamma$ coefficient being 100--1000 times larger than in
usual metals. However, this  $\gamma$ coefficient itself is strongly
$T$ dependent and rapidly decreases to ``normal'' values at higher
$T$. Similarly, the static magnetic susceptibility $\chi(T)$ crosses
over from  a Curie law $\chi(T) \sim 1/T$ at high temperatures to a
Pauli-type behavior for very low $T$: it approaches a finite value
$\chi(T=0)$,  which is also strongly enhanced compared to the Pauli
susceptibility of normal metals; the ratio between $\gamma$ and $\chi$
(Wilson ratio) is of order $1$, indicating that both enhancements are
caused by the same physical processes.

The heavy-fermion behavior is driven by the local moments of incompletely
filled $f$ shells of the lanthanide or actinide ions. At high temperature,
the weakly coupled $f$ electrons cause mainly incoherent  and
with decreasing temperature logarithmically growing spin-flip
scattering for the conduction electrons. Below a characteristic
temperature scale, a crossover to a coherent low-temperature phase is
observed. The $f$ electrons contribute significantly 
to the formation of heavy quasiparticles while their moments are
dynamically screened. Some heavy-fermion compounds such
as SmB$_6$ or Ce$_3$Bi$_4$Pt$_3$ exhibit similar behavior at high
temperature,  but for low $T$ a crossover to an insulating
heavy-fermion ground state with a narrow gap at the Fermi energy is
found.\cite{Hundley90} These heavy-fermion systems are termed
\emph{Kondo insulators} even though the insulating behavior is driven
by Fermi volume effects. In this case, the strong correlation induces
only a narrowing of the band gap rather than the insulator transition
itself as in  Mott-Hubbard insulators.

One particularly fascinating aspect of heavy-fermion compounds is that
despite the existence of local moments they can undergo a superconducting
phase transition, as observed\cite{Steglich79} in, e.g., CeCu$_2$Si$_2$.
In other systems magnetic phase transitions are reported, with a large
variety of types of order depending on the composition.\cite{Grewe91,Beyermann91}
In the weak-coupling or local-moment regime, the local $f$ moments tend
to order antiferromagnetically, while in the Kondo regime  itinerant
magnetism with an incommensurate ordering vector $\vec{Q}$ and strongly
reduced ordered moments is observed.\cite{SteglichGreweWelslau92}
This crossover from local to itinerant magnetism has been studied in
CeCu$_{(1-x)2}$Ge$_2$ by variation of the coupling constant upon doping
and has been attributed to the strongly temperature- and
coupling-dependent residual quasiparticle-quasiparticle interaction.
\cite{GreweWelslau88,GreweKeiterPruschke88,SteglichGreweWelslau92,Anders99}

In this paper, we focus on the calculation of transport properties
of heavy-fermion systems in the paramagnetic phase. Characteristic
information on heavy-fermion systems is obtained by measurements of 
the transport coefficients.
The Seebeck coefficient $S(T)$, or the thermoelectric power, is often
nonmonotonic, can exhibit different extrema, and even sign changes
\cite{steglich.77,schneider.83, jaccard.85,jaccard.90,ocko.99,Buehler-Paschen}
related to particle-hole asymmetries. Much of the recent interest in
the heavy-fermion thermoelectricity is stirred by its large narrow
peak at low temperatures which might be useful for solid-state
cooling devices.\cite{mahan.97} 

For high $T$, the resistivity $\rho(T)$ is determined by a
negative temperature coefficient, and  one usually observes a 
logarithmic, ``Kondo''-like increase of the resistivity $\rho(T)$ with
decreasing $T$. In Kondo insulators, $\rho(T)$ crosses over from low
values to an activation behavior  for smaller $T$, reflecting the
insulating ground state. In  metallic heavy-fermion systems---for example,
CePd$_3$ (Ref.~\onlinecite{Scoboria79}), CeAl$_3$ (Ref.~\onlinecite{Ott.75.84}),
or CeCu$_6$ (Ref.~\onlinecite{Onuki87})---a maximum is observed in $\rho(T)$
before the resistivity approaches a small residual value for $T\to 0$. At
low temperatures $T$,  $\rho(T)$  often obeys a $T^2$ law in such
materials. The logarithmic increase of $\rho(T)$  above a  characteristic
low-energy scale $T_\textrm{low}$
(for a discussion of low-energy scales see Sec.~\ref{sec:low-temperature})
is related to growing spin-flip scattering of the conduction electrons with
decreasing temperature as a  manifestation of the Kondo effect. Since the
resistivity must vanish for $T\to 0$ in a translational-invariant system with
a nondegenerate ground state, a maximum in $\rho(T)$ must connect these two regimes.
The observation of a $T^2$ behavior well below the characteristic temperature
scale $T_\textrm{low}$ in combination with a strongly enhanced $\gamma$ coefficient
of the specific heat indicates the formation of a Landau Fermi-liquid by
the heavy quasiparticles.

We show in a comprehensive study that the low-temperature
transport properties of heavy fermions can be understood in terms of a
minimalistic model of interaction between local and itinerate degrees of
freedom within the dynamical mean-field theory.\cite{Pruschke95,Georges96}
Spin-flip scattering between these degrees of
freedom yields a logarithmically increasing resistivity with
decreasing temperature. Our calculations reproduce the experimentally
observed maximum and the $T^2$ behavior at low temperatures
characteristic of coherent transport in the Fermi-liquid phase. The 
calculated optical conductivity exhibits a very narrow Drude peak
and an optical excitation gap characteristic of heavy-fermion
materials.\cite{DegiorgiAndersGruner2001}  The midinfrared peak is
located in the correct frequency range but with a too narrow width
compared to the experiment\cite{DegiorgiAndersGruner2001} due to the
lack of local crystal electric field excitations in our model.

We find very large absolute values for the thermoelectric power exceeding
$150\;\mu\mathrm{V/K}$ close to the Kondo
insulator regime at temperatures of the order of $T_\textrm{low}$. In
this case, the purely electronic figure of merit can be larger than 1.
The absolute magnitude of the thermoelectric power agrees very well
with the typical experimental reported values.
\cite{steglich.77,schneider.83,jaccard.85,jaccard.90,ocko.99,Buehler-Paschen} 
The sign changes of $S(T)$ depend very sensitively on the particle-hole
asymmetry and the band filling. Also in the experiment, the details of
the thermoelectric power vary strongly between different materials.
\cite{steglich.77,schneider.83,jaccard.85,jaccard.90,ocko.99,Buehler-Paschen}

The experimental evidence \cite{Maple95,Stewart01} compiled over the
past ten years also indicates  that even for heavy-fermion systems  with a
paramagnetic ground state, the temperature dependence of the specific
heat and the magnetic susceptibilities often do not agree with the
predictions of Fermi-liquid theory.\cite{Loehneysen96} Therefore, the
phenomenological term ``non-Fermi-liquid'' was attributed to such
regimes appearing in a large variety of different materials.
\cite{Maple95,Stewart01}  Despite a tremendous experimental and
theoretical effort it is, however, still not clear whether the non-Fermi-liquid
effects observed in heavy-fermion compounds are related to novel low-lying
nonlocal excitations in concentrated systems, true local non-Fermi-liquid
physics \cite{Millis93,Sachdev2001} or simply competing local energy
scales. \cite{AndersPruschke2006}

The physics in heavy-fermion compounds is driven by the interaction between
two distinct subsystems: localized, strongly correlated $f$ electrons
hybridizing with  extended conduction bands. The periodic Anderson model
(PAM) takes these ingredients into account, comprising of spin degenerate
conduction electrons, a lattice of correlated localized $f$ electrons and a
hybridization (cf.\ Sec.~\ref{sec:model}).
The conduction electrons experience spin-flip scattering from the magnetic
moments of the $f$ shells. On the other hand, the model also accounts for
the RKKY interaction between two localized magnetic moments in the particle-hole
channel,\cite{GreweKeiterPruschke88,GreweWelslau88} leading to the competition
between the Kondo screening and magnetic ordering.\cite{Doniach77} While
these qualitative aspects have been known for almost 30 years, a reliable
calculation of the full phase diagram of the PAM in more than one dimension
has not been possible yet.

In this  work, we apply the dynamical mean-field
theory\cite{Pruschke95,Georges96} (DMFT) mapping the PAM on an
effective single impurity Anderson model\cite{Anderson61} (SIAM),
which becomes exact\cite{Jarrell92,GeorgesKotliar92} in the limit of infinite
spatial dimension\cite{MetznerVollhardt89} for lattice models of
correlated electron systems. As impurity solvers for the effective
SIAM we use Wilson's numerical renormalization group\cite{Wilson75}
(NRG) for all interaction strength and the modified perturbation
treatment\cite{MartinRodero86,Potthoff97} (MPT) for small values of $U$. 
The NRG provides an accurate and non-perturbative
description of the effective site\cite{PruschkeBullaJarrell2000}
and yields the correct low-energy scale $T_K$ for the SIAM
(cf.\ Sec.~\ref{sec:diss-energy-scales}):
$T_K\propto\exp(-1/g)$, with the dimensionless coupling $g=J\rho_0(0)$,
the spin-flip scattering rate $J$, and $\rho_0(0)$ being the density of
states of the noninteracting conduction band of width $D$ at its band center.
Furthermore, the NRG provides detailed information on the possible
fixed-point structure of the effective site.
Compared to other possible and frequently used DMFT impurity solvers
such as quantum Monte Carlo\cite{HirschFye86} (QMC), exact
diagonalization (ED), and
noncrossing-approximation\cite{Grewe83,Kuramoto83,Bickers87} (NCA),
the NRG and MPT have the advantage to be applicable also for
very low temperatures.
This seems to be essential for the reproduction
of Fermi-liquid behavior and simultaneously identifying the relevant
low-temperature scale.

The transport properties are calculated in the linear response regime
using Kubo formulas. They relate the quantities of interest---e.g., the
frequency-dependent conductivity---to the charge and heat
current-current response functions, which are two-particle Green
functions. These response functions greatly simplify in any local
approximation since current-operator vertex corrections
vanish\cite{LorekAndersGrewe91} even in the presence of crystal field
levels:\cite{LorekAndersGrewe91,AndersCox97}  only the free
particle-hole propagator enters. From the $f$- and band-electron 
self-energies calculated within the DMFT-NRG and DMFT-MPT, the
static conductivity is obtained from the
transport integrals in the limit $\omega\to 0$, as well as
the thermoelectric power $S(T)$ and the electronic part of the
thermal conductivity $\kappa(T)$.
Recently, a very comprehensive study of the transport properties in
heavy-fermion systems as well as Kondo insulators using the local-moment
approach (LMA) has been published in a series of papers by Logan and
collaborators.\cite{LoganVid2003,VidLogan2004,LoganVid2005}
Previously, Costi and Manini\cite{Costi2002} investigated the
low-energy scales and temperature-dependent photoemission in the
$S=1/2$ Kondo lattice model using the DMFT-NRG.

The scope of our paper is to calculate the resistivity, optical
conductivity, the thermoelectric power, and the figure of merit for
arbitrary interaction strength and band filling within a single
approach, the DMFT-NRG. We address the question of the existence of
several low-temperature scales in the periodic Anderson model which
might manifest themselves in the transport properties.
The quality of the DMFT-NRG calculations is critically examined by a
comparison with the modified perturbation theory in the weak-coupling
regime ($U\to 0$), accurately described by the DMFT-MPT.

The paper is organized as follows: In Sec.~\ref{sec:theory} we
introduce the model and the notations as well as the DMFT and the two
impurity solvers used: the NRG and MPT.
In order to gain physical insight into the temperature
evolution of the transport properties, we discuss the
single-particle properties such as  the temperature-dependent
$f$-electron spectral function and self-energy of the PAM in
Sec.~\ref{sec:single-particle-dynamics}, since they
directly determine the transport properties through the Kubo formulas.
Section~\ref{sec:transport} is devoted to the transport theory for
the PAM and states explicitly equations for the different
transport coefficients investigated. We report results for the
transport quantities in Sec.~\ref{sec:transport-results}. The
temperature dependences of the resistivity $\rho(T)$, the
thermoelectric power $S(T)$, the thermal conductivity $\kappa(T)$, and
the temperature and frequency dependence of the optical conductivity
$\sigma(\omega,T)$ are shown for different values of $U$ and band
filling for metallic heavy-fermion systems as well as Kondo
insulators. In Sec.~\ref{sec:low-temperature} we define precisely
our different low-energy scales and discuss how they are linked. We
conclude with a summary and outlook in Sec.~\ref{sec:conclusion}.

\section{Theory}
\label{sec:theory}

\subsection{Model}
\label{sec:model}

As mentioned above, the Hamiltonian of the simplest version of the
periodic Anderson model is given by
\begin{align}
	\label{eq:pam-hamil}
	\hat{H} &= \sum_{\ks} \e_{\ks} c^\dagger_{\ks} c_{\ks}
	+\sum_{i\sigma} \e_{f\sigma} \hat{n}_{i\sigma}^f
	+\frac{U}{2}\sum_{i\sigma} \hat{n}^f_{i\sigma} \hat{n}^f_{i-\sigma} 
	\nonumber \\
	&\quad+V\sum_{i\sigma}(f^\dagger_{i\sigma} c_{i\sigma} +
	c^\dagger_{i\sigma}f_{i\sigma}).
\end{align}
Here,
$c_{\ks}$ ($c^\dagger_{\ks}$) destroys (creates) a conduction electron 
with spin $\sigma$, momentum $\k$ and energy $\e_{\ks}$. The energy
$\e_{f\sigma}$ denotes the spin-dependent single particle $f$-level
energy at lattice site $i$, $\hat{n}^f_{i\sigma} =
f^\dagger_{i\sigma}f_{i\sigma}$ is the $f$-electron
occupation operator (per site and spin),
$f_{i\sigma}$ ($f^\dagger_{i\sigma}$) destroys (creates) an $f$ electron 
with spin $\sigma$ at site $i$,
and $U$ denotes the on-site Coulomb repulsion between two $f$ electrons on 
the same site $i$. The uncorrelated conduction electrons
hybridize locally with the $f$ electrons via the matrix element $V$.
While Eq.~(\ref{eq:pam-hamil}) includes possible Zeeman splitting of
the energies in an external magnetic field $H$, we set $H=0$
throughout the remainder of the paper and treat all properties as
spin degenerate.

Even though only a single effective $f$ level is considered, this
model is quite general. It describes any heavy-fermion system
with odd ground-state filling of the $f$ shell, for which in a strong
crystal field environment  the degenerate Hund's-rule ground state
may be reduced to an effective spin-degenerate Kramers's doublet. In
addition, charge fluctuations to even $f$ fillings leave the $f$ shell
in  crystal field singlets.
The Hamiltonian contains four energy scales. The interplay between
$\e_f$ and $U$ controls the average $f$ filling as well as the local
moment formation for large $U$ and negative $\e_f$. The Anderson
width $\Gamma_0=V^2\pi\rho_0(0)$ determines the charge fluctuation scale of
the $f$ electrons with $\rho_0(0)$ being the density of states of the
noninteracting conduction band of width $D$ at its band center.

The total filling per site, $n_\textrm{tot} = 
\sum_{\sigma}(\expect{\hat{n}^c_{i\sigma}} + \expect{\hat{n}^f_{i\sigma}})$,
is kept constant by a temperature-dependent chemical potential $\mu(T)$. We
absorb the energy shifts into the band center $\e_c$ of the conduction
band, $\ek = \e_c + \tilde\e_{\k}$, as well as the $f$ level $\e_f$.
For $n_\textrm{tot} = 2$ and $U=0$, the uncorrelated system is an
insulator at $T=0$, since
the lower of the two hybridized bands is completely filled. According to
Luttinger's theorem a finite $U$ of arbitrary strength does not change
the Fermi volume which includes the full first Brillouin zone. As long
as the ground state does not change symmetry due to a phase
transition, the system remains an insulator at arbitrarily large Coulomb
repulsion. Therefore, the nonmetallic ground state of
Kondo insulators is not correlation induced, but it is already present for
the noninteracting system and a consequence of Luttinger's theorem. For
nonintegral values of $n_\textrm{tot}$, the paramagnetic phase of the system
must be metallic.

In general, we can distinguish three different adiabatically connected
regimes for $U\gg \Gamma_0$ and $\e_f<0$. In the mixed valence regime,
$\abs{\e_f}/\Gamma_0\approx 1$, the system is dominated by charge
fluctuations yielding a nonintegral value of the $f$ filling $n_f<1$. In the
stable moment regime $\abs{\e_f}\gg\Gamma_0$, the $f$ electrons remain
strongly localized and form a stable local moment, which tends to
order antiferromagnetically due to the RKKY interaction mediated by
the conduction electrons. These two regimes are connected by the Kondo
regime  for moderate ratios $\abs{\e_f}/\Gamma_0$. The competition between
screening of the local moment due to the Kondo effect and the RKKY
interaction makes this crossover regime  the most interesting one since it
can lead to long-range magnetic order of the residual magnetic moments.
Spin-density-wave, metallic and superconducting ground states are observed
in heavy-fermion materials\cite{Grewe91} which are believed to be
described by the Kondo regime.

\subsection{Dynamical mean-field theory}

Setting aside exact solutions in one dimension\cite{1DKondo1997}
using the Bethe ansatz for the Kondo lattice model, to our knowledge no
exact analytical solution has been found for the model
(\ref{eq:pam-hamil}) with finite $U$. Therefore, one has to rely on
suitable approximations for the PAM.
An obvious first approximation is the assumption of a purely local,
site-diagonal (i.e., $\vec{k}$-independent) self-energy, which for the
PAM is even better justified than for other lattice models of correlated
electron systems, as the first corrections are at least of order $V^6$.
Within a local self-energy approximation the complicated lattice problem
can be mapped on an effective impurity problem; i.e., the PAM can be
mapped on an effective SIAM.\cite{Anderson61} Such a mapping was first
used already about 20 years ago in connection with applications of the NCA
(Refs.~\onlinecite{Grewe83,Kuramoto83,Bickers87} and \onlinecite{KeiterKimball71})
to the PAM.\cite{Kuramoto85,Kim87,Grewe87,GreweKeiterPruschke88,Kim90}
A  self-consistency condition accounts for the feedback due to
the propagation of electrons through the lattice. Metzner and Vollhardt
\cite{MetznerVollhardt89} and M\"uller-Hartmann
\cite{MuellerHartmann89} noticed that the local approximation becomes
exact in the limit of infinite spatial dimensions ($d\to\infty$).
The effective site can be viewed as correlated atomic problem within a
time-dependent external field\cite{BrandtMielsch89} or an effective
SIAM.\cite{Kuramoto85,Jarrell92,GeorgesKotliar92}
This defines the self-consistency condition of the DMFT which has been
subject of two reviews.\cite{Pruschke95,Georges96} Within weak-coupling
$U$-perturbation theory it could be
shown\cite{SchweitzerCzycholl90b91a} that a local,
$\vec{k}$-independent self-energy is a good approximation for
realistic dimension $d = 3$ as corrections due to intersite
contributions to the self-energy are negligibly small.
However, phase transitions remain mean field like in DMFT
(Refs.~\onlinecite{Anders99}, \onlinecite{Georges96}, and
\onlinecite{Anders2002}) since $\k$-dependent fluctuations are not
included in a local approximation.

The following exact relations for the conduction electron Green
function  $G_{\sigma}(\k,z)$ and the $f$-electron Green function
$F_{\sigma}(\k,z)$ can be obtained for the PAM (\ref{eq:pam-hamil}):
\begin{subequations}
\begin{align}
\label{eq:conduction-electrons}
  G_{\sigma}(\k,z) &=
  \left[z-\e_{\ks}- \frac{\abs{V}^2}{z-\e_{f\sigma}-\Sigma^f_{\sigma}(\k, z) }\right]^{-1}, \\
  F_{\sigma}(\k,z) &=
  \left[z-\e_{f\sigma}-\Sigma^f_{\sigma}(\k, z)-\frac{\abs{V}^2}{z-\e_{\ks}}\right]^{-1},
\end{align}
\end{subequations}
where $z$ is any complex energy off the real axis.
Within a local approximation such as the DMFT, the $\k$-dependent
$f$-electron self-energy $\Sigma^f_{\sigma}(\k, z)$ is replaced by a
$\k$-independent $\Sigma^f_{\sigma}(z)$.
From Eq.~(\ref{eq:conduction-electrons}), one defines a self-energy
of the conduction electrons via
\begin{equation}
\label{eq:self-energy-gc}
  \Sigma_{\sigma}^{c}(z) =
  \frac{\abs{V}^2}{ z -\e_{f\sigma} -\Sigma^f_{\sigma}(z)},
\end{equation}
which can include a simple $\k$ dependence through the hybridization
matrix elements $\abs{V}^2$, here taken as constant.
For such a local self-energy $\Sigma_{\sigma}^{c}(z)$, the
site-diagonal conduction-electron Green function $G_\sigma$ can be
written as a Hilbert transformation
\begin{equation}
\label{eqn:g-loc}
  G_\sigma(z) = \frac{1}{N} \sum_{\k}G_{\sigma}(\k,z) =
  D(z-\Sigma_{\sigma}^{c}(z)),
\end{equation}
defined for arbitrary complex argument $z$ as
\begin{equation}
\label{eqn:hilbert-tranform}
  D(z) = \int_{-\infty}^{\infty} d\e \frac{\rho_0(\e)}{z-\e},
\end{equation}
where $\rho_0(\w)$ is the density of states of the  noninteracting
conduction electrons.

The DMFT self-consistency condition states that the site-diagonal
matrix element of the $f$-electron Green function of the PAM must
be equal to $F_{\textrm{loc},\sigma}(z)$ of an effective site problem
\begin{align}
\label{eq:pam-scc}
  F_{\sigma}(z) &= \frac{1}{N} \sum_{\k}F_{\sigma}(\k,z) = F_{\textrm{loc},\sigma}(z),
\displaybreak[0]\\
\label{eq:f-gf-local}
  F_{\textrm{loc},\sigma}(z) &=  \frac{1}{z-\e_{f\sigma} -\Delta_{\sigma}(z)
    -\Sigma^f_{\sigma}(z)},
\end{align}
with the same local $f$-electron self-energy $\Sigma^f_{\sigma}(z)$
for the lattice and the effective site. This defines the
self-consistency condition for the functions $\Sigma^f_{\sigma}(z)$
and $\Delta_{\sigma}(z)$.

Given the Green functions $F_\sigma(z)$ and $G_\sigma(z)$, their
spectral functions determine the local occupation numbers
\begin{subequations}
\begin{align}
\label{eq:nf}
  n_{f} &= \frac{1}{\pi} \sum_\sigma \int_{-\infty}^\infty d\w \;
  f(\w-\mu) \im F_\sigma(\w-i 0^+), \\
\label{eq:nc}
  n_{c} &= \frac{1}{\pi} \sum_\sigma \int_{-\infty}^\infty d\w \;
  f(\w-\mu) \im G_\sigma(\w-i 0^+),
\end{align}
\end{subequations}
where $f(\w)$ denotes the Fermi function. Then, the total filling per
site is given by $n_\textrm{tot} =  n_f +n_c$. As a matter of convenience,
we will perform an integral transformation such that $\mu$ is absorbed
into $\e_f$ and the band center $\e_c$; all energies will be measured
with respect to $\mu$. For a given lattice filling $n_\textrm{tot}$, we
have to adjust $\mu$ in addition to fulfill Eq.~(\ref{eq:pam-scc}).

Before we will discuss the solution of the effective site, let us
briefly comment on the implications of the analytical form of the
conduction-electron self-energy (\ref{eq:self-energy-gc}).
For a Fermi liquid, the imaginary part of $\Sigma^f$ vanishes
quadratically close to the chemical potential for $T\to 0$---i.e.,
$\im\Sigma^f_{\sigma}(\w-i0^+) \propto \w^2$.
At particle-hole symmetry---i.e., $n_\textrm{tot}=2$ for a symmetric
$\rho_0(\w)$---$\Sigma_{\sigma}^{c}(z)$ diverges like $1/z$ leading to
an insulator. Away from particle-hole symmetry, the denominator remains
finite and the imaginary part also must have Fermi-liquid properties
$\im\Sigma_{\sigma}^{c}(\w-i0^+) \propto \im\Sigma^f_{\sigma}(\w-i0^+)
\propto \w^2$. The real part is  very large, and therefore, the
spectral function of $G_\sigma(z)$ as well as $F_\sigma(z)$ samples
the high-energy band edges of $\rho_0(\w)$ yielding a hybridization
gap.
This analytic properties must be fulfilled by any approximate
solution of the  DMFT self-consistency condition (\ref{eq:pam-scc}). 

The low-temperature physics is determined by the temperature scale
$T_\textrm{low}$ which is only defined up to a constant factor.
We will use the renormalization of the Anderson width by the
quasiparticle spectral weight
\begin{equation}
   \label{eq:qp-weight-t0}
   T_0 = \Gamma_0\left[1- \left. \frac{\partial
         \re\Sigma^f(\w)}{\partial \w}
     \right|_{\w,T\to 0}
   \right]^{-1}
\end{equation}
as our choice of such a low-temperature scale $T_0\propto
T_\textrm{low}$ for our numerical analysis.\cite{VidLogan2004}
It is  related to the mass enhancement $m^*/m = \Gamma_0/T_0$.
We discuss other definitions of such a scale as well as
the possibility of several low-temperature scales
later in Sec.~\ref{sec:low-temperature}.

\subsection{Numerical renormalization group}
\label{sec:NRG}

The Green function $F_{\textrm{loc},\sigma}$, Eq.~(\ref{eq:f-gf-local}),
of the effective impurity problem can  be viewed as
the $f$-Green function of an effective
SIAM,\cite{Kim87,Jarrell92,GeorgesKotliar92,Pruschke95,Georges96}
\pagebreak[0]
\begin{align}
  \label{eq:effective-siam}
  \hat{H}_\textrm{eff}
  &= \sum_\sigma (\e_{f\sigma} -\mu) f^\dagger_\sigma f_\sigma
  + U \hat n_\uparrow \hat n_\downarrow
  + \sum_\sigma \int d\e\; (\e-\mu)
    d^\dagger_{\sigma\e}  d_{\sigma\e} \nonumber \\
  &\quad + \sum_\sigma \int d\e\; V\sqrt{\rho_\textrm{eff}(\e)}
    \left(
    d^\dagger_{\sigma\e} f_\sigma + f^\dagger_\sigma d_{\sigma\e} 
    \right),
\end{align}
with an energy-dependent hybridization function $\Gamma(\e) =
\pi V^2\rho_\textrm{eff}(\e) =\im\Delta(\e-i0^+)$ describing the
coupling of the $f$ electron to a fictitious bath of
``conduction electrons'' created by $d^\dagger_{\sigma\e}$
with density of states (DOS) $\rho_\textrm{eff}(\e)$.
The hybridization strength $V$ is chosen to be constant, defined via
\begin{equation}
  \pi V^2 = \int d\e \, \Gamma(\e),
\end{equation}
and equal to $V$ in the original model (\ref{eq:pam-hamil}).

We accurately solve the Hamiltonian (\ref{eq:effective-siam}) using Wilson's
NRG.\cite{Wilson75,KrishWilWilson80} The key ingredient in the NRG is a
logarithmic discretization of the continuous bath, controlled by the
parameter\cite{Wilson75} $\Lambda > 1$. The Hamiltonian is mapped onto a
semi-infinite chain, where the $N$th link represents an exponentially
decreasing energy scale $D_N \sim \Lambda^{-N/2}$. Using this
hierarchy of scales the sequence of finite-size Hamiltonians 
${\cal H}_N$ for the $N$-site chain is solved iteratively, truncating the
high-energy states at each step to maintain a manageable number of
states. The reduced basis set of ${\cal H}_N$ thus obtained is
expected to faithfully describe the spectrum of the full Hamiltonian
on the scale of $D_N$, corresponding\cite{Wilson75} to a temperature 
$T_N \sim D_N$ from which all thermodynamic expectation values are
calculated. The energy-dependent hybridization
function $\Delta(z)$ determines the coefficients of the semi-infinite
chain.\cite{BullaPruschkeHewson97}

The NRG is used to calculate the spectral function  of the Green
function  at finite temperatures.  We used a slight modification of
the algorithm\cite{AndersCzycholl2004} for finite temperature Green
functions by Bulla \etal\cite{BullaCostiVollhardt01} with the
broadening function\cite{BullaHewsonPruschke98}
\begin{equation*}
\delta(\w-E) \leadsto
 e^{-b^2/4} e^{-[\log(\w/E)/b]^2}/(\sqrt{\pi} b\abs{E}),
\end{equation*}
where we choose  $b=0.6$. As usual, the raw NRG Green functions
determine the self-energy $\Sigma^f_\sigma(z)$ by the exact ratio
\begin{equation}
\label{eqn:sigma-nrg}
  \Sigma^f_\sigma(z) = U \frac{M^\textrm{NRG}_\sigma(z)}{F_\textrm{loc}^\textrm{NRG}(z)},
\end{equation}
derived via equation of motion technique,\cite{BullaHewsonPruschke98} where
$M^\textrm{NRG}_\sigma(z) = 
\correlation{f_\sigma f^\dagger_{-\sigma} f_{-\sigma}}{f^\dagger_{\sigma}}(z)$.

\subsection{Modified perturbation theory}

The MPT is an approximation for calculating the self-energy of the SIAM
starting from the following ansatz:
\begin{equation}
	\Sigma^f_\sigma(z) = U n^{f}_{-\sigma} +
		\frac{\Sigma_{\sigma}^{(2)}(z)}
		{1-\beta_{\sigma}\; \Sigma_{\sigma}^{(2)}(z)}.
\end{equation}
It is based on second-order perturbation theory\cite{SchweitzerCzycholl89a}
(SOPT) relative to Hartree-Fock solution with $U$ as expansion parameter.
Here, $\Sigma_{\sigma}^{(2)}$ is the second-order contribution to the
self-energy (choosing $\beta_{\sigma}=0$ reproduces the SOPT),
\begin{multline}
	\Sigma^{(2)}_{\sigma}(z) = U^{2} \iiint
	\frac{\rho^\textrm{HF}_{\sigma}(\w_{1}) \rho^\textrm{HF}_{-\sigma}(\w_{2})
	\rho^\textrm{HF}_{-\sigma}(\w_{3})} {z-\w_{1}+\w_{2}-\w_{3}} \\
		\times \bigl\{f(\w_{1})[1-f(\w_{2})]f(\w_{3}) \\
		+ [1-f(\w_{1})]f(\w_{2})[1-f(\w_{3})]\bigr\}
		\:d\w_{1}\:d\w_{2}\:d\w_{3},
\end{multline}
with the Hartree-Fock $f$-electron spectral functions of the effective SIAM:
\begin{equation}
\rho^\textrm{HF}_{\sigma}(\w) = - \frac{1}{\pi}
\im\frac{1}{\w{+}i0^+ +\tilde{\mu} -\e_{f\sigma} -U n_{-\sigma}^f -
\Delta_{\sigma}(\w{+}i0^+)}.
\end{equation}
The parameter $\beta_{\sigma}$ is constructed such that the exactly solvable 
atomic limit $V=0$ as well as the first four spectral moments are correctly 
reproduced.\cite{MartinRodero86,Potthoff97} The  Hartree-Fock
$f$ occupation must equal the full $f$ occupation $n_{-\sigma}^f$
determining the effective Hartree-Fock chemical potential $\tilde{\mu}$.

\section{Results for the single-particle dynamics}
\label{sec:single-particle-dynamics}

All transport calculations rely on the results for the single-particle
Green functions of the PAM. We use a Gaussian model density of states for the
unperturbed conduction-electron system---i.e.,
$\rho_0(\w) = \exp[-(\w/t^*)^2/2]/(t^*\sqrt{2\pi})$---which is
appropriate for a $d$-dimensional hypercubic lattice in the limit
$d\to\infty$.\cite{MetznerVollhardt89} In the following, we measure energies
in units of $\Gamma_0 = \pi V^2 \rho_0(0)$---i.e., the hybridization is fixed at
$V^2= 2t^*\Gamma_0/\sqrt{2\pi}$---and choose $\sqrt2 t^* = 10\Gamma_0$.
As impurity solvers, we use the NRG as well as the MPT. The chemical potential
$\mu$ has to be determined self-consistently for a given total number of electrons,
$n_\textrm{tot}$, per site; in the following figures, energies (frequencies)  are
measured relative to the chemical potential $\mu$---i.e., we have $\mu = 0$
and the band center $\e_c$ is shifted accordingly.

\subsection{\textit{f}-electron spectral function and
self-energy at finite temperature \textit{T}}
\label{sec:f-spectral-function}

Using the NRG the calculation of the spectral functions for finite $T$
is numerically very challenging. In contrast to the  spectral
functions\cite{BullaHewsonPruschke98,PruschkeBullaJarrell2000} at $T=0$, where
only the ground-state excitations are relevant,  all excitations
contribute according to their statistical weight, but the NRG only
provides spectral information up to an energy scale $\w_N\propto \Lambda^{-N/2}$.
While the number of NRG iterations, $N$, is in principal arbitrary and
arbitrary small excitations could be resolved, Wilson has already pointed
out\cite{Wilson75} that the lowest energy scale
$\w_N$ should be identified with the temperature $T$ for which
thermodynamical expectation values are calculated. It
is obvious from the Lehmann representation of the spectral function that
eigenstates of the Hamiltonian with eigenenergies $E_M< T$ contribute
equally to the spectral functions. Therefore, we stop the NRG
iteration when $\w_N\approx T$ and must interpolate for frequencies
$\abs{\w}<\w_N$. The technical details of our finite-temperature algorithm
are described in the Appendix of Ref.~\onlinecite{AndersCzycholl2004}; other
possible approaches are found in Refs.~\onlinecite{BullaCostiVollhardt01} and
\onlinecite{CostiHewsonZlatic94,WeichselbaumDelft2006,PetersPruschkeAnders2006}.

\begin{figure}
  \includegraphics[width=0.8\columnwidth]{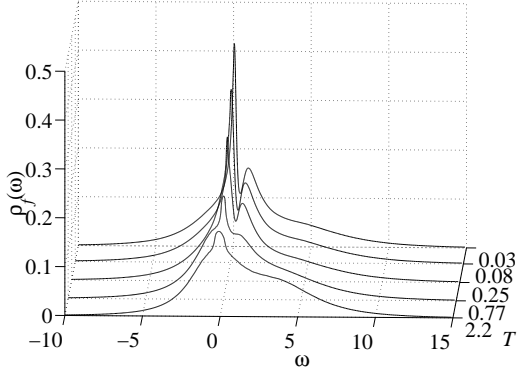}
  \caption{$f$-DOS $\rho_f(\w)$ calculated with DMFT-NRG for $U/\Gamma_0=5$, 
    $\e_f-\e_c=-2.5\Gamma_0$, chemical potential $\mu=0$, and a filling
    $n_\textrm{tot}=1.6$ at several finite temperatures $T/\Gamma_0$.
    NRG parameters: number of retained NRG states, $N_s=1500$,
    $\Lambda=1.6$, $\delta/\Gamma_0=10^{-3}$.}
  \label{fig:several-finite-T-f-dos}
\end{figure}

An exemplary series of $f$-spectral functions $\rho_f(\w)$ calculated
with DMFT-NRG for different temperatures $T$ and fixed $U=5\Gamma_0$
is shown in Fig.~\ref{fig:several-finite-T-f-dos} for the metallic
regime  with a filling $n_\textrm{tot} = 1.6$ and
$\e_f-\e_c=-2.5\Gamma_0$. A hybridization pseudogap develops close
to the chemical potential $\mu=0$ for  $T$  decreasing below a
characteristic temperature $T_\textrm{low}$. The moderate value of $U$
leads to a spectrum where the high-energy charge excitations are not
well separated from the resonance close to the chemical potential.

\begin{figure}
  \includegraphics[width=0.8\columnwidth]{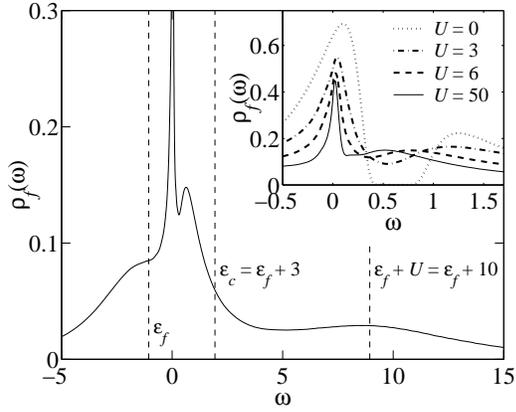}
  \caption{$f$-DOS $\rho_f(\w)$ calculated with DMFT-NRG for $U/\Gamma_0=10$,
    $\e_f-\e_c=-3\Gamma_0$, chemical potential $\mu=0$, filling $n_\textrm{tot}=1.6$,
    and finite temperature $T=0.0003\Gamma_0$.
    A three-peak structure with peaks at $\e_f$, $\mu$, and $\e_f+U$ is visible.
    In the inset the transition from a hybridization gap for $U=0$ (exact result)
    to a pseudogap for $U/\Gamma_0\in\{3,6,50\}$ is shown.
    NRG parameters: as in Fig.~\ref{fig:several-finite-T-f-dos}.}
  \label{fig:finite-T-f-dos}
\end{figure}

The typical structure of the $f$-spectral function $\rho_f(\w)$ calculated
with DMFT-NRG for a larger value of  $U/\Gamma_0=10$, $\e_f-\e_c=-3\Gamma_0$,
at a temperature $T=0.0003\Gamma_0$ well below $T_{\textrm{low}}$ is depicted
in Fig.~\ref{fig:finite-T-f-dos} (a detailed discussion of possible
estimates for $T_{\textrm{low}}$ is found in Sec.~\ref{sec:low-temperature}).
A  pronounced peak structure with a pseudogap  dominates the
low-energy part of the spectrum in the vicinity of the chemical
potential similar to the one shown in Fig.~\ref{fig:several-finite-T-f-dos}.
In addition, we observe two shallow high-energy peaks, one at $\e_f$
below $\mu$, one at $\e_f + U$ which corresponds to double
occupancy of the $f$ levels. It is always correctly positioned by the
NRG, independent of the value of $U$, but with a linewidth too large
due to the NRG broadening procedure (see Refs.~\onlinecite{AndersCzycholl2004}
and \onlinecite{BullaHewsonPruschke98} for details).

In the inset of Fig.~\ref{fig:finite-T-f-dos} the development of the
hybridization gap of the $f$-spectral function $\rho_f(\w)$ with
increasing $U>0$ is shown and compared with the exact result for
$U=0$. For $U=0$, we obtain from an uncorrelated hybridized band a
hybridization gap above $\mu=0$ of a width $\propto V$. It arises from
the divergence of the real part of $\Sigma_{\sigma}^{c}(z\to \e_f)$ given in
Eq.~(\ref{eq:self-energy-gc}): the high-energy part of the free density of
states $\rho_0$ defines the shape of the gap
via the exact relation between $G_\sigma$ and $F_\sigma$,
\begin{equation}
  F_\sigma(z) = \frac{\Sigma^c_\sigma(z)}{V^2}\bigl\{1 + \Sigma^c_\sigma(z)
     D[z-\Sigma_{\sigma}^{c}(z)] \bigr\}.
\end{equation}
For a DOS $\rho_0(\w)$ of Gaussian shape the hybridization gap is,
strictly speaking, only a pseudogap even at $U=0$. A real gap only
arises for densities of states  with well-defined sharp band edges such
as a semielliptical or a true finite-$d$ DOS.
However, with increasing $U$ the gap will always evolve into a pseudogap
anyway due to the finite lifetime of the quasiparticles, reflected in the
growing imaginary part of $\Sigma_\sigma^f(\omega)$. This effect is clearly
seen in the inset to Fig.~\ref{fig:finite-T-f-dos}.
The width of the pseudogap and of the whole Kondo
resonance peak structure is decreasing with increasing $U$ stemming from the
the reduction of the quasiparticle weight at the chemical potential.

\begin{figure}
  \includegraphics[width=0.8\columnwidth]{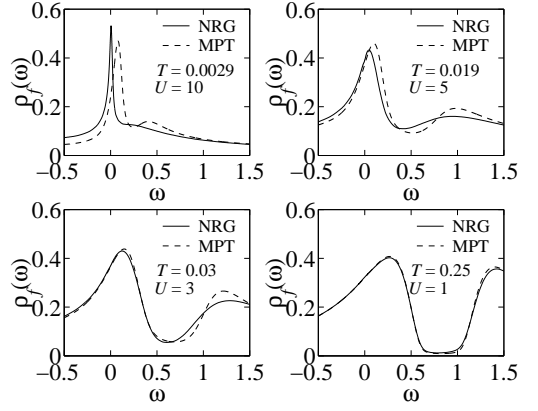}
  \caption{Comparison between the $f$-DOS $\rho_f(\w)$ calculated with DMFT-NRG
    and DMFT-MPT for $U/\Gamma_0\in\{10,5,3,1\}$ and $\e_f-\e_c=-U/2$
    at finite temperatures $T/\Gamma_0$ as indicated.
    Here, the chemical potential is $\mu=0$ with filling $n_\textrm{tot}=1.6$.
    NRG parameters: as in Fig.~\ref{fig:several-finite-T-f-dos}.}
  \label{fig:finite-T-f-dos-comparison}
\end{figure}

A comparison of the $f$-spectral functions for finite $T$ obtained in DMFT-NRG
and DMFT-MPT for $\e_f-\e_c=-U/2$ and $n_\textrm{tot} = 1.6$ (metallic case)
and four different values of $U$ is displayed in Fig.~\ref{fig:finite-T-f-dos-comparison}.
Again, only a pseudogap is found at finite $U$ which is narrowing with increasing $U$.
As expected, the DMFT-MPT curves agree very well with the DMFT-NRG
graphs for small $U$. This is not surprising, because the MPT is based on
$U$-perturbation theory and must, therefore, become correct for sufficiently
small $U$.  Already for $U/\Gamma_0=5$, however, one observes  deviations between
the MPT and NRG approaches. For $U/\Gamma_0=10$ ($U/\sqrt{2}t^*=1$) the width of the 
resonance peak at the Fermi energy is obviously too large in MPT; i.e., quantitatively 
there is a strong overestimation of the low-energy scale within
DMFT-MPT. For large $U$ the MPT cannot reproduce the correct Kondo
temperature scale, as the physics in this regime is driven by the spin-flip scattering
$J$  proportional to $1/U$ and is, therefore, nonperturbative in $U$.

\begin{figure}
  \includegraphics[width=0.8\columnwidth]{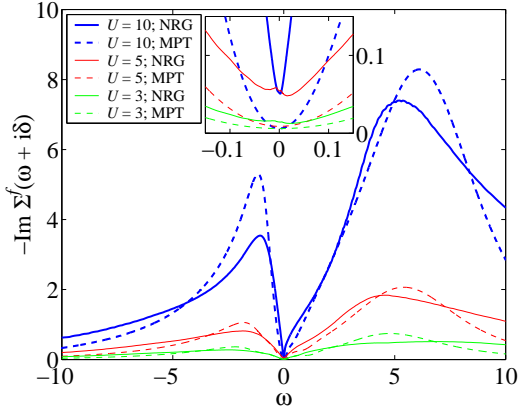}
  \caption{(Color online) Comparison between the self-energy $\Sigma^f(\w)$ in
    DMFT-NRG and DMFT-MPT for $U/\Gamma_0\in\{3,5,10$\}. The inset
    shows a close-up of the vicinity of $\mu=0$. 
    All parameters are as in Fig.~\ref{fig:finite-T-f-dos-comparison}, in particular
    finite temperatures $T/\Gamma_0\in\{0.03,0.019,0.0029\}$.}
  \label{fig:finite-T-sigma-f}
\end{figure}

Figure \ref{fig:finite-T-sigma-f} shows the frequency dependence of the
(imaginary part of the) $f$-electron self-energy obtained in DMFT-NRG and
DMFT-MPT, corresponding to Fig.~\ref{fig:finite-T-f-dos-comparison}.
Here, the imaginary part
of the self-energy is finite even at $\w=0$, and away from the chemical
potential one observes a quadratic $(\w-\mu)^2$ behavior as expected for
Fermi liquids. The finite value at $\w = \mu = 0$ is not only due to the
finite $T$, but it has also a contribution from a finite imaginary part
$\delta$ (i.e., an additional Lorentzian broadening), which we had to
introduce for numerical reasons. As we observed numerical instabilities in
the self-consistency equations (\ref{eq:pam-scc}) and (\ref{eq:f-gf-local})
as well as inaccuracies in the local self-energy
(\ref{eqn:sigma-nrg}) in NRG, we solved the set of equations 
(\ref{eq:pam-scc}) and (\ref{eq:f-gf-local}) in the complex plain
at a finite shift $\delta/\Gamma_0=10^{-3}$ away from the real axis to
obtain stable numerical solutions. Physically this artificial broadening
$\delta$ can be interpreted to simulate the effects of impurity scattering,
yielding a finite lifetime corresponding to a finite imaginary part of an
additional ``disorder'' self-energy. But one also sees from 
Fig.~\ref{fig:finite-T-sigma-f} that---in spite of the same small
imaginary part $\delta$---the MPT result for $\im\Sigma^f(0)$ is
systematically smaller than the NRG result, even for small $U$. This is
an indication that the finite value in NRG is not only caused by the
temperature offset and $\delta$ but also by the additional Gaussian
broadening\cite{BullaHewsonPruschke98,AndersCzycholl2004} described in
Sec.~\ref{sec:NRG} and additional
numerical errors stemming from the relation (\ref{eqn:sigma-nrg}) to
determine the self-energy in NRG at finite temperature.

\subsection{Temperature dependence of band-electron self-energy \texorpdfstring{$\Sigma^c(\w)$}{}}
\label{sec:c-self-energy}

Since the conduction-electron self-energy $\Sigma^c(\w,T)$ strongly 
influences the transport properties of the PAM, we study its 
imaginary part at $\w=0$---i.e., the conduction-electron scattering rate.
According to Eq.~(\ref{eq:self-energy-gc}), it is given by
\begin{equation}
  \im\Sigma_{\sigma}^{c}(0) = \frac{\abs{V}^2 \im\Sigma^f_{\sigma}(0)}{
  [\e_{f\sigma} +\re\Sigma^f_{\sigma}(0) ]^2 + [\im\Sigma^f_{\sigma}(0)]^2}.
  \label{eq:im-sigma-c}
\end{equation}
Analytically, we can distinguish two cases.

(i) If $\abs{\e_{f\sigma}+\re\Sigma^f_{\sigma}(0)} \ll \abs{\im\Sigma^f_{\sigma}(0)}$,
the conduction-electron scattering rate is reciprocal proportional to the
$f$-electron scattering rate---i.e.,
$\im\Sigma_{\sigma}^{c}(0)\approx V^2/\im\Sigma^f_{\sigma}(0)$. This is the
case for the Kondo insulator regime, where
$\e_{f\sigma} + \re\Sigma^f_{\sigma}(0)\to 0$.

(ii) Only when
$\abs{\e_{f\sigma}+\re\Sigma^f_{\sigma}(0)} \gg \abs{\im\Sigma^f_{\sigma}(0)}$
is the denominator of Eq.~ (\ref{eq:im-sigma-c}) dominated by the real part
$[\e_{f\sigma}+\re\Sigma^f_{\sigma}(0) ]^2$ and the
$c$-scattering rate becomes proportional to $\abs{\im\Sigma^f_{\sigma}(0)}$.

\begin{figure}
  \includegraphics[width=0.8\columnwidth]{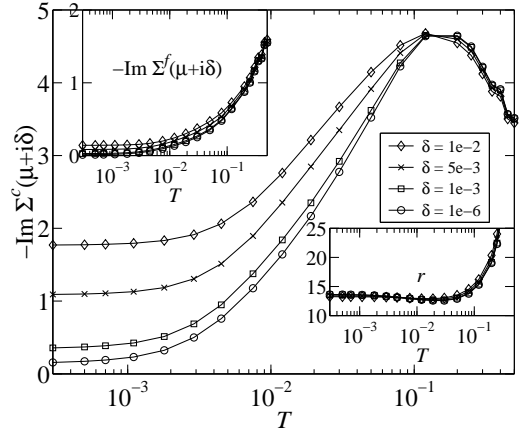}
  \caption{Influence of the shift $\delta$ on $\im\Sigma^c(\mu,T)$
    for $U/\Gamma_0=10$, $\e_f-\e_c=-U/2$, and $n_\textrm{tot}=1.6$.
    The ratio $r=V^2/[\e_{f} +\re\Sigma^f(0)]^2$ (right inset) is
    almost constant for low $T$, here $r\approx14\gg1$.
    This leads to the enhancement of $\im\Sigma^c(\mu,T)$ for low $T$
    compared to $\im\Sigma^f(\mu,T)$ as shown in the left inset.
    NRG parameters as in Fig.~\ref{fig:several-finite-T-f-dos}.}
  \label{fig:finite-T-sigma-c-U=10} 
\end{figure}

We show the temperature dependence of the $c$-electron scattering rate
and its dependence on $\delta$ for a fixed value of $U/\Gamma_0=10$ in
Fig.~\ref{fig:finite-T-sigma-c-U=10}. 
While the $f$-electron scattering rate remains very small for low $T$
(cf.\ left inset of Fig.~\ref{fig:finite-T-sigma-c-U=10}), the 
$c$-electron scattering rate has a much larger finite $T\to 0$
value. This originates from the small but finite
imaginary part of the conduction-electron self-energy
(\ref{eq:im-sigma-c})  as a consequence of the small (artificial) imaginary part
$\delta$, here $\delta/\Gamma_0\in\{10^{-2},5\times10^{-3},10^{-3},10^{-6}\}$ and
the numerical error in the ratio of two Hilbert transformed 
spectral functions in Eq.~(\ref{eqn:sigma-nrg}).
The small finite $\abs{\im\Sigma^f_{\sigma}(0)}$ is enhanced if the ratio 
$r:=V^2/[\e_{f\sigma} +\re\Sigma^f_{\sigma}(0) ]^2$ is larger than $1$,
as it is usually the case since the hybridization gap is very close to the
chemical potential, and $-\e_{f\sigma} -\re\Sigma^f_{\sigma}(0)$ is an
estimate for its location.  As one can see from the right inset of
Fig.~\ref{fig:finite-T-sigma-c-U=10}, we indeed have a ratio $r\gtrsim
14>1$. Therefore, we can 
estimate the $c$-electron scattering rate by
\begin{equation*}
  \im\Sigma_{\sigma}^{c}(0) \sim 14 \im\Sigma^f_{\sigma}(0)
\end{equation*}
for the chosen parameters $U/\Gamma_0=10$ and $n_{\textrm{tot}}=1.6$,
which explains why the artificial finite imaginary part is even more 
important and pronounced for $\im\Sigma_{\sigma}^{c}(0)$
than it is for  $\im\Sigma_{\sigma}^{f}(0)$.

\subsection{Renormalized band structure}

\begin{figure}
  \includegraphics[width=0.9\columnwidth]{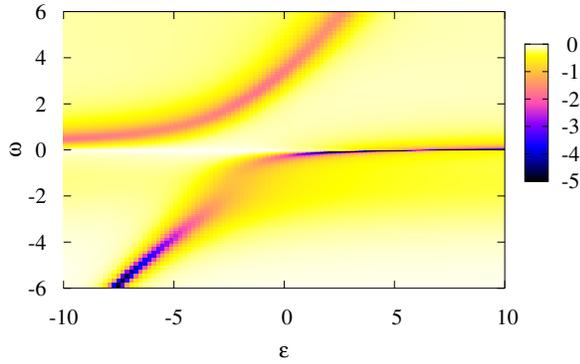}
  \caption{(Color online) Contour plot of the total energy-dependent
    density of state $\rho(\e,\w)=\rho_f(\e,\w)+\rho_c(\e,\w)$ for
    $U/\Gamma_0=8$, $\e_f-\e_c=-U/2$, $n_\textrm{tot}=1.6$, $\mu=0$,
    $T/\Gamma_0=3\times 10^{-4}$.
    NRG parameters: as in Fig.~\ref{fig:several-finite-T-f-dos}.}
  \label{fig:renormalized-bands}
\end{figure}

In the previous section, Sec.~\ref{sec:f-spectral-function}, we presented
the local, or $\k$-summed spectral function. In the Fermi-liquid phase
of the model, new quasiparticles are formed as a mixture of $f$ and $c$
degrees of freedom. The resulting renormalized band structure can be
visualized by plotting the energy $\e=\e_{\k}$ and frequency dependent
density of states $\rho(\e,\w)=\rho_f(\e,\w)+\rho_c(\e,\w)$, where
\begin{subequations}
\begin{align}
	\rho_f(\e,\w) &= \im F_\sigma(\e,\w-i0^+)/\pi, \\
	\rho_c(\e,\w) &= \im G_\sigma(\e,\w-i0^+)/\pi,
\end{align}
\end{subequations}
as a two-dimensional (2D) color (online) contour plot displayed in
Fig.~\ref{fig:renormalized-bands}. For $n_\textrm{tot}=1.6$, the chemical
potential $\mu=0$ lies at the top of the lower hybridized band. For $V=0$, we
would have a sharp line at $\w=\e-\e_c$ with purely $c$ character, while
two lines with fractional weight of $1/2$ at $\w=\e_f=\e_c-U/2$ and
$\w=\e_f+U=\e_c+U/2$ with $f$ character could be found. At finite $V$, the
$f$ electrons become part of the Fermi volume. Indeed, the almost
dispersionless and therefore heavy quasiparticles close to the chemical
potential have mainly $f$ character. The second band is located above
the chemical potential. We note---not shown here---that this renormalized
band picture remains valid even well above the low-temperature
scale. The dark colored peaks of $\rho(\e,\w)$ are broadened by 
the increasing imaginary part of the self-energy but can still be
traced by the zeros of the real part of the reciprocal Green
functions $G_\sigma(\e,\w)$ and $F_\sigma(\e,\w)$. This observation
turns out the be very important to understand the midinfrared peak in
the optical conductivity.\cite{Degiorgi99,DegiorgiAndersGruner2001}

\section{Transport theory}
\label{sec:transport}

To describe the electronic transport within the PAM we start from the
standard relations for the generalized transport coefficients, according to
which  the electrical current density $\vec{J}$ and the heat current density
$\vec{q}$ depend linearly on the electric field $\vec{E}$ and the temperature
gradient $\nabla T$:
\begin{subequations}
\begin{align}
	\vec{J} &= L_{11} \vec{E} + L_{12} \bigl(-\tfrac{1}{T}\nabla T\bigr), \\
	\vec{q} &= L_{21} \vec{E} + L_{22} \bigl(-\tfrac{1}{T}\nabla T\bigr).
\end{align}
\end{subequations}
All coefficients are calculated within the linear response approach,
starting from similar Kubo formulas.\cite{Luttinger64,Mahan81} For
symmetry reasons, $L_{12}=L_{21}$ holds.

For example, the real part of the frequency dependent (optical) conductivity
tensor\cite{Voruganti92,Mahan81,Czycholl2000}
$\sigma(\w)=L_{11}(\w)$ is related to the current-current
correlation function and written as
\begin{equation}
	\sigma_{\alpha\beta}(\w) 
	=-\frac{1}{\w NV_0} \im\correlation{j_{\alpha}}{j_{\beta}^\dagger}(\w+i0^+),
\end{equation}
where $V_0 = a^3$ is the volume of the unit cell and $N$ counts the number
of lattice sites.
It has been shown\cite{CzychollLeder81} that the current operator of the PAM
has two contributions: a conduction-electron part and a part proportional to
$\nabla V_{\k}$. The $f$ electrons do not appear in the current, since they
do not disperse. For a $\k$-independent hybridization, only the conduction
electrons carry the electrical and heat currents:
\begin{equation}
  \vec{j} = e\sum_{\ks} \vec{v}_{\k} c^\dagger_{\ks} c_{\ks},
\end{equation}
where $\vec{v}_{\k} = \frac{1}{\hbar}\nabla_{\k}\ek$ is the group velocity.
Hence, the current-susceptibility tensor $\correlation{\vec{j}}{\vec{j}^\dagger}(z)$
is connected to the particle-hole Green function 
\begin{equation}
 \correlation{\vec{j}}{\vec{j}^\dagger}(z)
= e^2\sum_{\sigma\sigma'\k\kk} \vec{v}_{\k} \vec{v}_{\kk}^T 
\correlation{c^\dagger_{\ks} c_{\ks}}{c^\dagger_{\kk\sigma'} c_{\kk\sigma'}}(z).
\end{equation}
In a cubic crystal, the  conductivity is isotropic: $\sigma_{\alpha\beta}(\w)
= \sigma(\w)\openone$.
From now on, we will consider only the $xx$-component of the conductivity
$\sigma(\w) \equiv \sigma_{xx}(\w)$.

In general, the full two-particle Green function
$\correlation{c^\dagger_{\ks} c_{\ks}}{c^\dagger_{\kk\sigma'} c_{\kk\sigma'}}(z)$ 
involves  vertex corrections which reflect residual particle-particle
interactions.\cite{Mahan81} However, in the limit $d\to\infty$ it was shown that
current operator vertex corrections vanish.\cite{Khurana90,SchweitzerCzycholl91b}
Thus, it is consistent with the DMFT assumption of a $\k$-independent self-energy
that these vertex corrections vanish for any lattice model of correlated electron
systems. For the special case of a local approximation for the PAM this was
already shown in Refs.~\onlinecite{LorekAndersGrewe91}, \onlinecite{AndersCox97},
and \onlinecite{CoxGrewe88}, as for symmetry reasons
\begin{equation*}
 \sum_{\k} \vec{v}_{\k} \abs{V_{\k}}^2 G_{\k}(z+\w)G_{\k}(z) = 0.
\end{equation*}
Therefore, we obtain
\begin{align}
 \correlation{j_x}{j_x^\dagger}(\w+i0^+)
 &= \frac{e^2}{\hbar^2} \sum_{\ks}
\left(\frac{\partial\ek}{\partial k_x} \right)^2
\int_{-\infty}^\infty d\w' f(\w') \rho_c(\ek,\w') \nonumber \\
&\quad\times[G_\sigma(\k,\w'+\w + i0^+) \nonumber \\
&\quad+ G_\sigma(\k,\w'-\w - i0^+)].
\label{eq:-chi-jj-v}
\end{align}
Within the DMFT, the lattice one-particle Green function depends only on the
(complex) energy $z$ and bare band dispersion $\ek$: $G_{\sigma}(\k,z) =
G_\sigma(\ek,z)$. Then 
\begin{equation}
 \frac{1}{N}\sum_{\k} \left(\frac{\partial\ek}{\partial k_x} \right)^2
 A(\ek) = \int_{-\infty}^\infty d\e \,\tilde\rho_0(\e) A(\e),
\end{equation}
with
\begin{equation}
  \tilde\rho_0(\e) = \frac{1}{N}\sum_{\k}
  \left(\frac{\partial\ek}{\partial k_x} \right)^2
  \delta(\e -\ek).
\label{eq:rho-tilde}
\end{equation}
$\tilde\rho_0(\e)$ has been evaluated approximately in large
dimensions\cite{Pruschke93} as
\begin{equation}
  \tilde\rho_0(\e) = \frac{(at^*)^2}{d} \rho_0(\e) + O(1/d^2)
\end{equation}
on a hyper-cubic lattice. Then, Eq.~(\ref{eq:-chi-jj-v}) can be reduced to a
sum of Hilbert transforms, defined in Eq.~(\ref{eqn:hilbert-tranform}).

By taking the limit $\w\to 0$ in Eq.~(\ref{eq:-chi-jj-v}),
the static conductivity $\sigma=L_{11}$ is obtained, and we
get for the generalized transport coefficients:
\begin{subequations}
\label{eq:lin-resp}
\begin{align}
	\label{eq:cond}
	L_{11} &= \frac{e^2}{\hbar a}
		\int_{-\infty}^\infty[-f'(\w)] \tau(\w) d\w, \\
	L_{12} &= \frac{e}{\hbar a}
		\int_{-\infty}^\infty[-f'(\w)] (\w-\mu) \tau(\w) d\w, \\
	L_{22} &= \frac{1}{\hbar a}
		\int_{-\infty}^\infty[-f'(\w)] (\w-\mu)^2 \tau(\w) d\w.
\end{align}
\end{subequations}
Here $\tau(\w)$ represents a generalized relaxation time defined as
\begin{equation}
	\tau(\w) = \frac{2\pi}{d}(t^*)^2\int_{-\infty}^\infty \rho_0(\e) \rho^2_c(\e,\w) d\e
\label{equ:relaxation-time}
\end{equation}
and $f'(\w)$ is the derivative of the Fermi function.
Only in the Fermi-liquid regime does one have
$\tau(\w)\propto 1/\im\Sigma_c(\w+i0^+)$ and the linearized Boltzmann
transport theory is recovered.

The thermal conductivity $\kappa$ and the thermoelectric power $S$
are then given by\cite{Czycholl2000,Mahan81}
\begin{align}
	S &= \frac{1}{T}\frac{L_{12}}{L_{11}}
		= \frac{k_B}{\abs{e}} \frac{\abs{e}L_{12}}{k_B T L_{11}},
\label{eq:lin-res-thermo}\\
	\kappa &= \frac{1}{T}\left(L_{22}-\frac{L_{12}^2}{L_{11}}\right).
\end{align}

The thermoelectric power $S$ is defined as the proportionality constant
between an applied  temperature gradient and the measured voltage drop
in the absence of a current flow. The Peltier coefficient given by the
ratio of heat and electrical current is related to the thermoelectric
power by $\Pi=T S$.
Note that $\abs{e}L_{12}/k_B T L_{11}$ is dimensionless and
$k_B/\abs{e}\approx 86\;\mu\mathrm{V/K}$. Therefore, the thermoelectric
power is given in absolute units; only the scale of temperature axis must
be fixed by experiment.

If we assume one electron per unit cell of the volume $a^3$,
$a=10^{-10}\;\mathrm{m}$, the resistivity $\rho=\sigma^{-1}$ has the natural unit
\begin{equation}
  \label{eq:sigma_0}
  \sigma_0^{-1} = \hbar a/e^2 \approx 41\;\mu\Omega\;\mathrm{cm}.
\end{equation}
Similarly, if we assume $\Gamma_0=100\;\mathrm{meV}$,
the thermal conductivity is given in units of
\begin{equation}
  \frac{k_B\Gamma_0}{\hbar a} \approx 0.21\;\mathrm{W/(K\;cm)}.
\end{equation}

Finally, the dimensionless figure of merit is defined as 
\begin{equation}
	ZT = \frac{T\sigma S^2}{\kappa},
\end{equation}
which measures the efficiency of a thermoelectric material.

\section{Results for the transport properties}
\label{sec:transport-results}

\subsection{Resistivity}

The (static or direct current) resistivity $\rho(T)$ is obtained from the
reciprocal of $\sigma(T)$, Eq.~(\ref{eq:cond}). To achieve better convergence
of the DMFT equation at low temperatures again a finite offset $\delta>0$ from
the real axis was introduced. As discussed in Sec.~\ref{sec:f-spectral-function},
the NRG provides accurate spectral functions only for frequencies $\w\ge T$.
In the linear-response transport-integrals (\ref{eq:lin-resp}), however, only
the low-frequency spectral information in the Fermi window $\abs{\w}<2T$
contributes. This intrinsic property of the NRG makes it very difficult
to obtain reliable spectral information for low frequencies at finite
temperatures needed for calculating transport properties.

\begin{figure}
  \includegraphics[width=0.8\columnwidth]{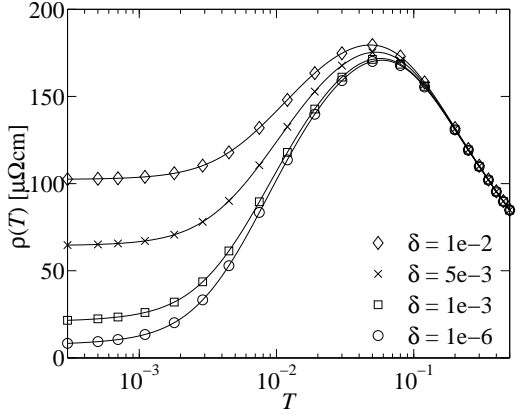}
  \caption{Influence of the shift $\delta$ on the static
    resistivity in DMFT-NRG for $U/\Gamma_0=10$, $\e_f-\e_c=-U/2$,
    and $n_\textrm{tot}=1.6$.
    All parameters are as in the corresponding
    Fig.~\ref{fig:finite-T-sigma-c-U=10}.}
  \label{fig:rho-t-U10-delta}
\end{figure}

In Fig.~\ref{fig:rho-t-U10-delta}, the resistivity is displayed for fixed
filling $n_\textrm{tot}=1.6$, $U/\Gamma_0=10$, $\sqrt2 t^*=10\Gamma_0$ and
various values of the shift $\delta$ as in Fig.~\ref{fig:finite-T-sigma-c-U=10}.
Physically, a finite $\delta$ can be interpreted as simulating the effects of
lattice defects and/or of a small but finite concentration $c$ of nonmagnetic
impurities giving rise to a finite ``impurity self-energy'' imaginary part
(inverse scattering time)  $\delta \sim c$,
which leads to a finite residual resistivity for $T\to 0$. Figure
\ref{fig:rho-t-U10-delta} shows that the residual resistivity $\rho(T=0)$ is,
in fact, increasing with increasing $\delta$. As discussed already in
Sec.~\ref{sec:c-self-energy}, the finite
imaginary part of the conduction-electron self-energy (\ref{eq:im-sigma-c})
is much enhanced compared to the imaginary part of the $f$-electron
self-energy, which explains the strong dependence on $\delta$ seen in
Fig.~\ref{fig:rho-t-U10-delta}. However, even in the limit $\delta\to 0$
a finite $\rho(T=0)$ is obtained for $T\to 0$ in DMFT-NRG.
Not only the finite $\delta$ but also the additional Gaussian broadening
and the limited accuracy of the NRG in the regime $\w < T$  causes the
finite self-energy imaginary part and the finite $\rho(0)$. We reproduce
the typical metallic HFS behavior\cite{Scoboria79, Ott.75.84, Onuki87}
within our DMFT-NRG treatment:
a resistivity increasing with increasing $T$ for low $T$, a maximum of the
order $100\;\mu\Omega\;\mathrm{cm}$ at a characteristic temperature
$T_\textrm{max}$ (about $10$--$200\;\mathrm{K}$), and a $\rho(T)$
(logarithmically) decreasing with increasing $T$ for $T > T_\textrm{max}$.

\begin{figure}
  \includegraphics[width=0.8\columnwidth]{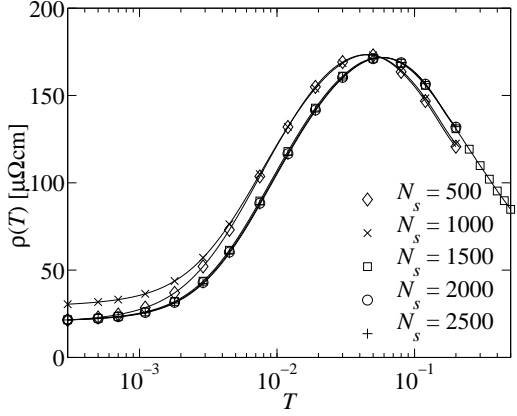}
  \caption{Resistivity $\rho(T)$ obtained in DMFT-NRG for different
    number of kept states $N_s$ and fixed values of
    $\delta/\Gamma_0 = 10^{-3}$ and $\Lambda =1.6$.
    Other parameters as in Fig.~\ref{fig:rho-t-U10-delta}.}
  \label{fig:rho-t-U10-vs-N}
\end{figure}

For the same parameters as in Fig.~\ref{fig:rho-t-U10-delta} the dependence
of $\rho(T)$ on the NRG parameter $N_s$ (number of states kept) is
shown in Fig.~\ref{fig:rho-t-U10-vs-N}. One observes that for $N_s \ge 1500$,
$\rho(T)$ no longer depends on $N_s$. Therefore, all further DMFT-NRG
calculations have been performed for this $N_s=1500$.

\begin{figure}
  \includegraphics[width=0.8\columnwidth]{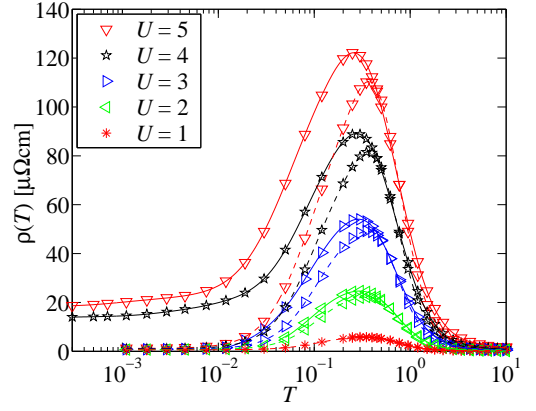}
  \caption{(Color online) Comparison between the resistivity $\rho(T)$ in
    DMFT-NRG (solid lines) and DMFT-MPT (dashed lines).
    for different $U$, $\e_f-\e_c=-U/2$, and fixed total
    filling $n_\textrm{tot} =1.6$ (metallic case).
    NRG Parameters: as in Fig.~\ref{fig:several-finite-T-f-dos}.}
  \label{fig:rho-t-fixed-n-tot-vs-U-MPT}
\end{figure}

For fixed $\delta/\Gamma_0 = 10^{-3}$, a total filling
$n_\textrm{tot}=1.6$, a difference between the band center $\e_c$ and
$\e_f$ of  $\e_f-\e_c=-U/2$ and several small values of $U$ the DMFT-NRG
and DMFT-MPT results for $\rho(T)$ are compared in
Fig.~\ref{fig:rho-t-fixed-n-tot-vs-U-MPT}. One observes that in the
high-temperature regime both results agree and that the overall agreement
becomes the better the smaller $U$ is, as expected already from our discussion
in Sec.~\ref{sec:f-spectral-function}, Fig.~\ref{fig:finite-T-f-dos-comparison}.
But for the same small $\delta$ the residual resistivity $\rho(0)$ in
DMFT-MPT is much smaller than in DMFT-NRG. In DMFT-MPT the finite $\rho(0)$
is solely determined by the finite $\delta$ whereas in DMFT-NRG
the additional broadening and numerical inaccuracies at very small $\w$
contribute.

\begin{figure}
  \includegraphics[width=0.8\columnwidth]{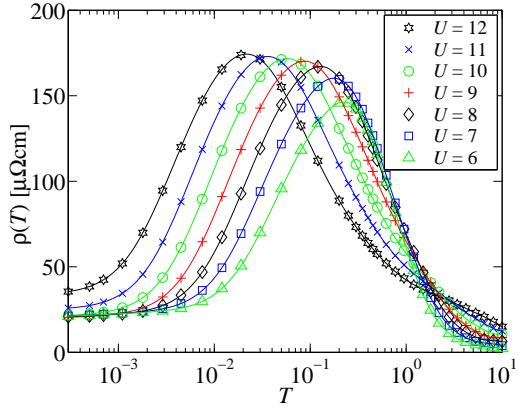}\\ (a)

  \includegraphics[width=0.8\columnwidth]{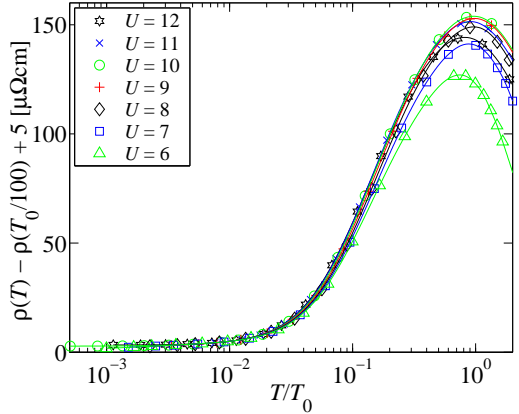}\\ (b)

  \caption{(Color online) Resistivity $\rho(T)$  as function of $T$ for different $U$,
    $\e_f-\e_c=-U/2$, and fixed total filling $n_\textrm{tot} =1.6$.
    The calculations were done in DMFT-NRG, thus continuing the NRG series
    of Fig.~\ref{fig:rho-t-fixed-n-tot-vs-U-MPT} to higher $U$.
    NRG Parameters: as in Fig.~\ref{fig:several-finite-T-f-dos}.
    In (b) the same data is plotted as
    $\rho(T)-\rho(T_0/100) +5\;\mu\Omega\;\mathrm{cm}$ versus $T/T_0$ with
    $T_0$ as in Eq.~(\ref{eq:qp-weight-t0}).}
  \label{fig:rho-t-fixed-n-tot-vs-U}
\end{figure}

For the same parameter set, but higher values of the correlation $U$,
DMFT-NRG resistivity results are plotted in
Fig.~\ref{fig:rho-t-fixed-n-tot-vs-U}. The temperature $T_\textrm{max}$,
at which the  resistivity has its maximum, is shifted to lower values for
increasing $U$; as discussed in Sec.~\ref{sec:low-temperature} one obtains
an exponential dependency of $T_\textrm{max}$ on $U$. For sufficiently
strong $U$ the peak height at $T_\textrm{max}$ is nearly $U$ independent.
This is seen once more from Fig.~\ref{fig:rho-t-fixed-n-tot-vs-U}(b), which
shows the scaling properties  plotting the resistivity $\rho(T)-\rho(T_0/100)$
versus $T/T_0$ [for the low-temperature scale $T_0$ see
Eq.~(\ref{eq:qp-weight-t0}) and Sec.~\ref{sec:low-temperature}].
While for $U/\Gamma_0<7$, where high- and low-temperature scales are not very
well separated, the maximum of the resistivity and the peak height show a $U$ 
dependence, we reach a universality regime for large $U$. Note that
$T_0$ is of the order of the position $T_\textrm{max}$ of the maximum of the
resistivity while the $T^2$ behavior of the resistivity is only
observed below the coherent scale $T_\textrm{cor}$ which is two orders of
magnitude  smaller.

\begin{figure}
  \includegraphics[width=0.8\columnwidth]{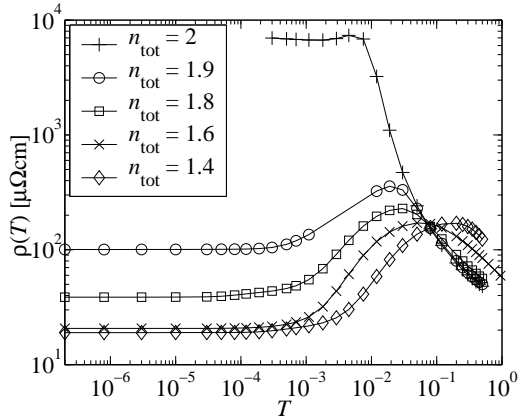}
  \caption{Resistivity $\rho(T)$  for different total fillings
    $n_\textrm{tot}$,   $U/\Gamma_0=10$,  $\e_f-\e_c=-U/2$,
    $\delta/\Gamma_0 = 10^{-3}$.
    NRG parameters: as in Fig.~\ref{fig:several-finite-T-f-dos}.}
  \label{fig:rho-t-U10-vs-n-tot}
\end{figure}

\begin{figure}
  \includegraphics[width=0.8\columnwidth]{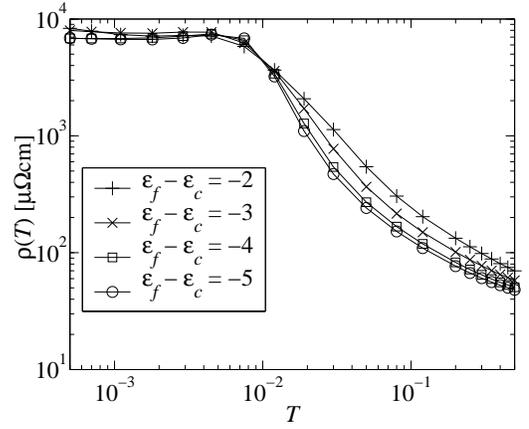}

  \caption{Resistivity $\rho(T)$  for different $\e_f-\e_c$,
    $\delta/\Gamma_0 = 10^{-3}$,   $U/\Gamma_0=10$, $n_\textrm{tot}=2$
    (Kondo insulators).
    NRG parameters: as in Fig.~\ref{fig:several-finite-T-f-dos}.}
  \label{fig:rho-t-U10-n-tot2-vs-ef}
\end{figure}

The dependence of the resistivity $\rho(T)$ on the total occupation
$n_\textrm{tot}$ is shown in Fig.~\ref{fig:rho-t-U10-vs-n-tot} for
fixed $U/\Gamma_0=10$ and $\e_f - \e_c = -U/2$.
We see that the (artificial) residual resistivity $\rho(0)$ increases
whereas the maximum temperature $T_\textrm{max}$ decreases when approaching
the Kondo insulator regime at $n_\textrm{tot}=2$ (symmetric PAM). In this
case $\rho(T)$ saturates for $T < 10^{-2}$, which is typical
for Kondo insulators.\cite{Hundley90} Experimentally, this is due to
impurities; in our calculations it is a consequence of the
finite-energy imaginary part $\delta$.
For fixed
$n_\textrm{tot}=2$ and different values of $\e_f-\e_c$, $\rho(T)$ is shown in
Fig. \ref{fig:rho-t-U10-n-tot2-vs-ef}; obviously, the Kondo insulator
behavior is not only obtained in the symmetric case but always for
$n_\textrm{tot}=2$ in accordance with the Luttinger theorem. The
(artificial) finite $\rho(0)$ depends hardly on $\e_f-\e_c$ but only on the
finite $\delta$.

\subsection{Optical conductivity}

\begin{figure}
  \includegraphics[width=0.8\columnwidth]{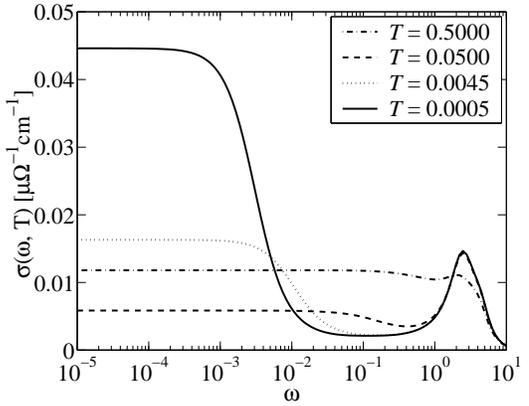}

  \caption{Optical conductivity for four different temperatures
    $T\approx0.01,0.1,1,10 T_\textrm{max}$, $n_\textrm{tot}=1.6$,
    and a fixed $U/\Gamma_0=10$, $\e_f-\e_c=-U/2$. 
    NRG parameters: as in Fig.~\ref{fig:several-finite-T-f-dos}.}
  \label{fig:opical-conductivity-U=10-n1-6}
\end{figure}

The optical conductivity $\sigma(\w)$ obtained in the metallic case for
$n_\textrm{tot}=1.6$ is shown in
Fig.~\ref{fig:opical-conductivity-U=10-n1-6} for different temperatures $T$. For
very low $T$ one observes a Drude peak at low frequencies and, in addition, a
``midinfrared'' peak at finite frequency $\w \sim \sqrt{t^* T_0}$. With
increasing $T$ the Drude peak quickly decreases (in accordance with the
strong increase of the static resistivity), whereas the midinfrared peak
remains nearly unchanged, and the remainders of the Drude peak and the
midinfrared peak merge into a broad structure in the high temperature
regime.

\begin{figure}
  \includegraphics[width=0.8\columnwidth]{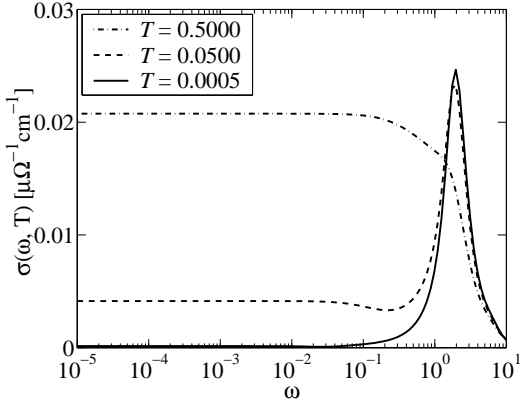}

  \caption{Optical conductivity for three different temperatures $T$,
    $n_\textrm{tot}=2$ (i.e., Kondo insulators), and a fixed
    $U/\Gamma_0=10$, $\e_f-\e_c=-U/2$.
    NRG parameters: as in Fig.~\ref{fig:several-finite-T-f-dos}.}
  \label{fig:opical-conductivity-U=10-n2}
\end{figure}

Figure \ref{fig:opical-conductivity-U=10-n2} shows  $\sigma(\w)$ for a Kondo
insulator, where for very low $T$ the Drude peak is absent 
and only the midinfrared peak is present due to interband transitions.
With increasing $T$,
however, the gap disappears, and the metallic heavy-fermion  behavior at
high temperatures is recovered.

This behavior can easily be understood in terms of  the renormalized bands
discussed in Sec.~\ref{sec:single-particle-dynamics} and shown
in Fig.~\ref{fig:renormalized-bands}.  With  the assumption of a
constant density of states $\rho_0(\mu)$, $\sigma(\w)$ can be approximated by
\begin{equation}
  \label{eq:opti-approx}
\frac{\sigma(\w)}{\sigma_0} = 
\frac{i\rho_0(\mu)}{2\w}
\int_{-\infty}^\infty
\frac{d\w'[f(\w') -f(\w'+\w)]}{\w +\Sigma^c(\w'-i\delta) -
\Sigma^c(\w'+\w + i \delta)}.
\end{equation}
The optical conductivity
$\sigma(\w)$ has a maximum when the real part  of the denominator on
the right-hand side of Eq.~(\ref{eq:opti-approx}) vanishes.\cite{Jarrell95}
This is obviously the case for $\w\to 0$ for metallic HFSs
in the Fermi-liquid regime, which
yields the Drude peak. An additional maximum is found in the vicinity
of the minimal intraband transition energy close to  the chemical
potential, which yields the midinfrared peak. An estimate for its position
is the smallest interband transition energy for $\vec{q}=0$ given by
$\Delta_\textrm{opt}\propto \sqrt{t^*T_0}$ which
fits perfectly with the observed midinfrared peak of a large variety
of different heavy-fermion compounds.\cite{DegiorgiAndersGruner2001}
Its temperature dependence is weak since the spectral weight moved from lower
to higher excitation energies is small.
On the other hand, the midinfrared peak width depicted in
Fig.\ \ref{fig:opical-conductivity-U=10-n1-6} and
reported in Ref.\ \onlinecite{Jarrell95} comes out much narrower than the
experimentally observed peaks.\cite{Degiorgi99} This is expected
because in real materials, all Hund's-rule multiplets contribute to
the scattering of the conduction electrons in the midinfrared region. Our
simplified model, however, only contains the lowest Hund's-rule
doublet. The total optical response is furnished by a superposition
of all scattering resonances  which leads to an effectively broad
resonance as observed experimentally.

\subsection{Thermoelectricity}

\begin{figure}
   \includegraphics[width=0.8\columnwidth]{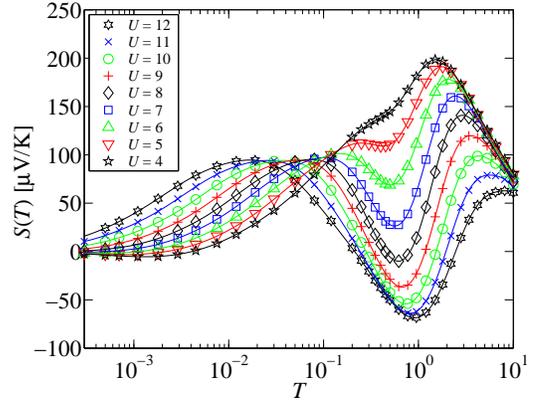}\\ (a)

   \includegraphics[width=0.8\columnwidth]{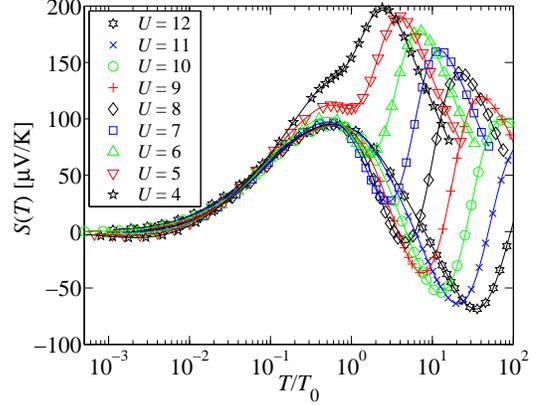}\\ (b)

   \caption{(Color online) Thermoelectric power for various values of $U$ and fixed
     total occupation $n_\textrm{tot}=1.6$ and $\e_f-\e_c=-U/2$. 
     Panel (b) shows the same date but on a rescaled axis $T/T_0$ with
     $T_0$ as in Eq.~(\ref{eq:qp-weight-t0}).
     All Parameters: as in Fig.\ \ref{fig:rho-t-fixed-n-tot-vs-U}. 
     }
   \label{fig:thermopower-vs-U-n1-6}
\end{figure}

The thermoelectric power $S(T)$ measures the ratio between electrical and
heat current divided by the temperature, and its sign is related to the
integrated particle-hole asymmetry relative to the chemical potential. In
Fig.~\ref{fig:thermopower-vs-U-n1-6}, $S(T)$ is plotted for
$\e_f-\e_c=-U/2$ and fixed total filling  for a variety of different
values of $U$ [for the low-temperature scale $T_0$ see
Eq.~(\ref{eq:qp-weight-t0}) and Sec.~\ref{sec:low-temperature}]. We obtain
very large absolute values for $S(T)$, of the magnitude $50\;\mu\mathrm{V/K}$
up to $150\;\mu\mathrm{V/K}$; note that the thermoelectric power is obtained
in absolute units, as already mentioned in Sec.~\ref{sec:transport}.
Similar to the resistivity, the thermoelectric power exhibits a low-temperature
peak which is correlated with the maximum of the resistivity which is an
analytical consequence of Eq.~(\ref{eq:lin-res-thermo}). In addition, we
observe a second extremum moving to higher temperatures with increasing $U$
which results from the charge fluctuations on the energy scale $\e_f-\mu$.

It is very interesting to note that for $U/\Gamma_0>8$ the thermoelectric
power changes sign in an intermediate-temperature regime.
Several extrema and similar sign changes were found experimentally
for several HFS materials, in particular below $80\;\mathrm{K}$ for
CeCu$_{2.2}$Si$_2$.\cite{Steglich96,AndersHuth2001} 
As we have discussed in Sec.~\ref{sec:single-particle-dynamics}, the
details of the spectrum close to the hybridization gap  depend on the
free single-particle DOS $\rho_0(\w)$. In addition, our model does not
include other Hund's-rule multiplets. Therefore, we only can make a
qualitative comparison with experiments. Remarkably, however, we find
a sign change of the thermoelectric power for large $U$ and a large
absolute value of $S(T)$ comparable to the experimental results.

\begin{figure}
  \includegraphics[width=0.8\columnwidth]{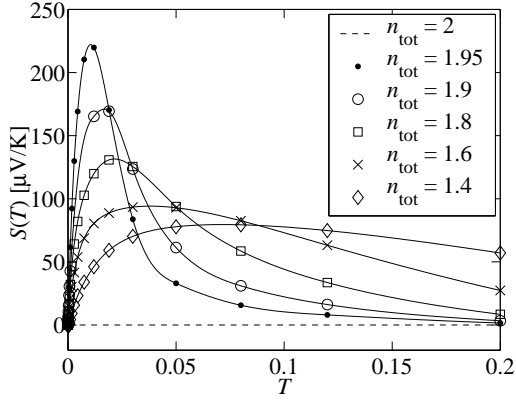}
  \caption{Thermoelectric power for a fixed  $U/\Gamma_0 =10$ and various
    occupations   $n_\textrm{tot}=1.4,1.6,1.9,1.95,2$ and $\e_f-\e_c=-U/2$.
    All parameters: as in Fig.~\ref{fig:rho-t-U10-vs-n-tot}.}
  \label{fig:thermopower-vs-n-tot-U10}
\end{figure}

\begin{figure}
  \includegraphics[width=0.8\columnwidth]{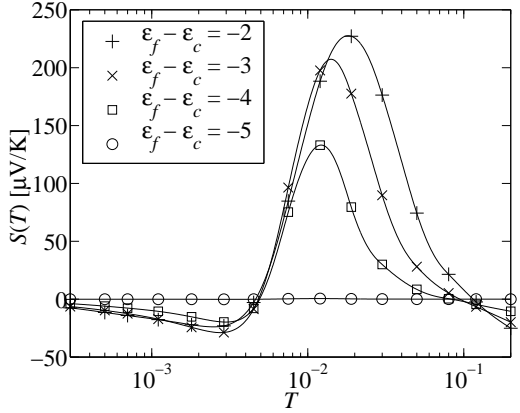}
  \caption{Thermoelectric power for $U/\Gamma_0 = 10$, $n_\textrm{tot}=2$ (Kondo
    insulator regime) and different values for $\e_f$ (or $\e_f - \e_c$).
    All parameters: as in Fig.~\ref{fig:rho-t-U10-n-tot2-vs-ef}.}
\label{fig:thermopower-vs-e-f-ntot2}
\end{figure}

The thermoelectric power $S(T)$ for different fillings $n_\textrm{tot}$,
$U/\Gamma_0=10$, and $\e_f - \e_c = -U/2$ is depicted in
Fig.\ \ref{fig:thermopower-vs-n-tot-U10}.
For $n_\textrm{tot}=2$ the thermoelectric power vanishes for all temperatures
because of particle-hole symmetry. Otherwise $S(T)$ exhibits a peak with
increasing height when approaching the Kondo insulator regime
$n_\textrm{tot}\to 2$. For the Kondo insulator situation $n_\textrm{tot}=2$
and various $\e_f - \e_c$ the thermoelectric power $S(T)$ is shown in
Fig.~\ref{fig:thermopower-vs-e-f-ntot2}. Only in the symmetric case
$\e_f = - U/2 = -5\Gamma_0$ does $S(T)$ vanish due to particle hole symmetry;
away from the symmetric case, the peaks reach extremely large absolute values
of $S$ up to $200\;\mu\mathrm{V/K}$. This shows that Kondo insulators could be
candidates for thermoelectric applications.

\begin{figure}
  \includegraphics[width=0.8\columnwidth]{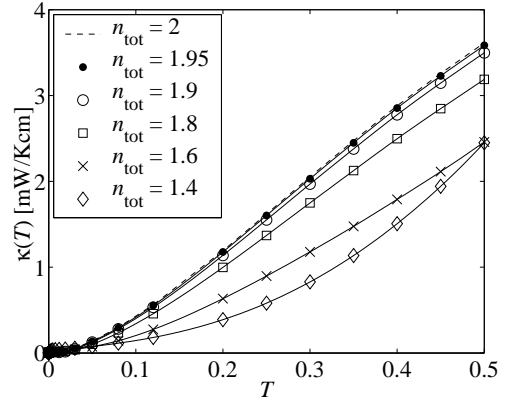}
  \caption{Electronic part of thermal conductivity for
    $U/\Gamma_0 =10$, $\e_f-\e_c=-U/2$.
    All parameters: as in Figs.~\ref{fig:rho-t-U10-vs-n-tot} and
    \ref{fig:thermopower-vs-n-tot-U10}.}
  \label{fig:thermal-conductivity-vs-n-tot-U10}
\end{figure}

\begin{figure}
  \includegraphics[width=0.8\columnwidth]{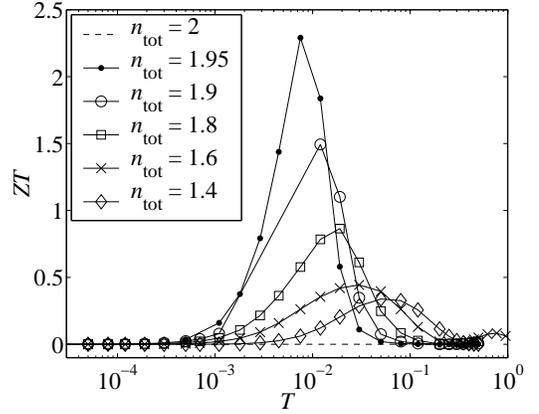}
  \caption{Figure of merit for $U/\Gamma_0 =10$, $\e_f-\e_c=-U/2$.
    All parameters: as in Figs.~\ref{fig:rho-t-U10-vs-n-tot} and
    \ref{fig:thermopower-vs-n-tot-U10}.}
  \label{fig:figure-of-merit-vs-n-tot-U10}
\end{figure}

However, the figure of merit $Z=S^2 \sigma/\kappa$, or the
dimensionless value $ZT$ and not only the Seebeck
coefficient $S(T)$ is of relevance for efficient thermoelectric
cooling, where $\kappa$ is the thermal
conductivity. Since approaching the Kondo insulator reduces the
conductivity $\sigma$ as well as shifting the  peak to very low
temperatures the nominator of $ZT$ will be rather small.
Fig.~\ref{fig:thermal-conductivity-vs-n-tot-U10} shows the electronic part
of the thermal conductivity $\kappa(T)$. As expected $\kappa(T)$ is rapidly
increasing with increasing $T$ and the overall behavior is similar as that
observed experimentally.\cite{Buehler-Paschen}

The resulting temperature dependence of the figure of merit is shown in
Fig.~\ref{fig:figure-of-merit-vs-n-tot-U10}. According to our calculations
a figure of merit larger than $1$ can be obtained in HFSs in certain
low-temperature regimes, showing that, in fact, they are candidates for
thermoelectric applications.

\section{Low-temperature scales}
\label{sec:low-temperature}

There has been a long debate whether low-temperature properties are
governed by  one or more low-energy scales in the paramagnetic phase of the
PAM.\cite{Nozieres,Jarrell95,
TahvildarJarPruFre1999,PruschkeBullaJarrell2000,LoganVid2005}
In the SIAM, the screening of the local moments occurs on the  scale
of Kondo temperature   $T_K$ where the system crosses over from the
local-moment to the strong-coupling fixed point when lowering the
temperature. But Nozi\`eres has argued 
that for the lattice with a macroscopic concentration of
$f$ moments there is an ``exhaustion'' of the conduction
electrons\cite{Nozieres,TahvildarJarPruFre1999}
available for screening of those $f$ moments.
However, it is questionable whether the concept of individually screened
independent single impurities still holds in a lattice, because one conduction
electron could rather contribute to the screening of several local moments.
In this context, it has been suggested that two different
low-temperature scales exist in the paramagnetic phase of the PAM:
namely, the single-impurity Kondo scale $T_K$ and a strongly reduced
lattice scale $T_\textrm{low} \ll T_K$ at which the screening of the local
moments happens.
It was argued that at $T_K$ only $n_\textrm{scr} = \rho_c(0) T_K$
conduction electrons are available to screen $n_f$ local $f$ electrons,
where $\rho_c(0)$ is the conduction-electron DOS at the chemical potential.
However, in Nozi\`eres's work\cite{Nozieres} only phenomenological estimates
and no precise mathematical definitions for different temperature
scales are given.

It is tempting to identify the possibly existing additional scale
$T_\textrm{low}$ with the ``coherence temperature''---i.e., the scale on
which the heavy quasiparticles are formed and/or on which the $T^2$ behavior
in the resistivity is observable.  Experimentally, however, no evidence
was found for such a large separation of energy scales in
heavy-fermion compounds.\cite{Grewe91,Stewart01} On the contrary,
substitution of Ce by La in Ce$_{1-x}$La$_x$Cu$_{6}$ exhibits nearly perfect
scaling of the Ce contribution to the specific heat\cite{Onuki87} from the
diluted impurity situation ($x \approx 1$) to the lattice case 
($x = 0$) indicating that there is only one relevant energy scale. Recent
theoretical investigations by Vidhyadhiraja and Logan gave evidence for such
a single low-temperature scale,\cite{LoganVid2003,VidLogan2004,LoganVid2005}
using Fermi-liquid assumptions for the $f$-electron self-energy. Their
arguments are consistent with the notion that a new thermodynamically
relevant energy scale can only occur as crossover scale from competing
low-temperature fixed points. Our NRG calculations for the effective
site, moreover, gave no evidence for new fixed points of the SIAM other
than those already discussed in Ref.\ \onlinecite{KrishWilWilson80}. In
fact, we always find a strong-coupling fixed point independent of the
filling connecting Kondo insulators adiabatically to metallic heavy-fermion
systems.

In the literature, different notions of what is implied by the term
``low-temperature scale''  have led to a controversy
which could have been resolved by stating  clearly the precise mathematical
definitions of such distinct scales.  The
low-temperature scale $T_\textrm{low}$ allows to map any
temperature-dependent property $P(T)$ over a limited low-temperature
range\cite{VidLogan2004} onto a corresponding dimensionless 
universal function $p(x=T/T_\textrm{low})$ where all details of the initial
microscopic Hamiltonian enter this single scale $T_\textrm{low}$. From a
renormalization group perspective, it describes the crossover from one
fixed point to the $T=0$ fixed point. If such a quantity $P$ has
distinct analytical behavior in different temperature intervals, it
is tempting to assign  additional low-temperature scales to identify
such different regimes. If, however, a universal function $p(x)$ can be found
such that $P(T)$  collapses onto this curve for all low-temperature
regimes, these distinct scales contain no additional physical information.
The confusion in the literature arises by ignoring that these
admittedly different low-temperature scales might ultimately be connected
and reflect the same physical phenomena. If, however, different
scaling regimes can be identified using two or more low-temperature
scales, it should  correspond to additional fixed points of the
Hamiltonian.

\subsection{Definition of different low-temperature scales} 
\label{sec:def-energy-scales}

\begin{figure}
  \includegraphics[width=0.8\columnwidth]{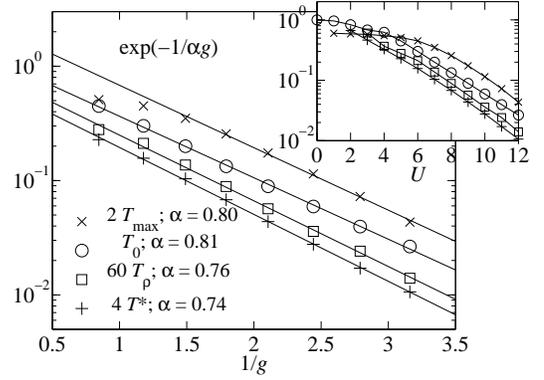}

  \caption{The four different low-temperature scales plotted versus
    the reciprocal dimensionless coupling constant $1/g$
    [cf.\ Eq.~(\ref{eq:coupling})]. The inset
    shows the same data as function of $U$.
    The stars mark the maximum of the resistivity
    $T_\textrm{max}/\Gamma_0$ the crosses the screening of the  spin moment
    of the effective site. The squares indicate
    $T_\rho=1/\sqrt{\sigma_0A}$, while the mass enhancement factor
    is plotted by the circles.
    Parameters: $\e_f-\e_c= -U/2$, $n_\textrm{tot}=1.6$,
    number of retained NRG states $N_s=1500$,
    $\Lambda=1.6$, $\delta/\Gamma_0=10^{-3}$.
  }

  \label{fig:energy-scales}
\end{figure}

From our data,  we have defined \emph{four different} possible
low-temperature scales characterizing \emph{different} physical
phenomena in order to identify more than one low-temperature
scale. These scales are
(i) the low-temperature scale\cite{VidLogan2004} $T_0=\Gamma_0 m/m^*$,
  Eq.~(\ref{eq:qp-weight-t0}),
  defined by the quasiparticle spectral weight,
  measuring the reciprocal mass enhancement 
  $m/m^*$ for $T\to 0$,
  \begin{equation*}
    m/m^* = \left[1- \tfrac{\partial \re\Sigma^f(\w)}{\partial \w}
    \Bigr|_{\w,T\to 0} \right]^{-1};
  \end{equation*}
(ii) the $A$ coefficient of the resistivity $\rho(T) = \rho_0 + AT^2$
  valid   in the coherent Fermi-liquid regime,
  $T_{\rho} = (\sigma_0 A)^{-1/2}$, where $\sigma_0^{-1}$ is
  the natural unit of the resistivity given in  Eq.\ (\ref{eq:sigma_0});
(iii) the position of the maximum of the resistivity curve $T_\textrm{max}$
  for metallic heavy-fermion systems;
and (iv) the screening temperature $T^*$ of the  impurity moment\cite{Wilson75}
  $\expect{S^2_z}_\textrm{imp}(T^*)=0.05$.

Other possible scales such as the width of
the hybridization gap are not very well defined since the position and
depth of the gap as well as the shape of the spectral function are
strongly dependent on the Fermi volume.
In the following, we will show that all scales are connected by a fixed
ratio when varying the interaction strength $U$ for fixed occupancy
$n_\textrm{tot}$ or Fermi volume.

The mass enhancement factor $m^*/m$ as well as the
$A$ coefficient of the resistivity contains information on the
low-lying excitations of the lattice Fermi-liquid fixed point as pointed
out by Logan and collaborators.\cite{VidLogan2004}
Both should be proportional to a low-temperature scale $T_\textrm{low}$.
We contrast this with two estimates of
a possible second low-temperature scale. It was argued that this
scale governs the crossover from a regime of incoherent scattering of
the conduction electrons to a coherent Fermi-liquid ground state of the
lattice.\cite{PruschkeBullaJarrell2000}
This truly new scale would mark the deviation of a
periodic Anderson model from the properties of the SIAM. We take
the position of the maximum of the resistivity $T_\textrm{max}$ as one
clear indicator of such a scale as well as the screening of the impurity
effective moments which defines the temperature $T^*$.
However, we like to point out that our definitions of the
low-temperature scales do not imply an energy hierarchy
$T_0 <T_\textrm{max}$.

\subsection{Discussion of the scales}
\label{sec:diss-energy-scales}

\begin{figure}
   \includegraphics[width=0.8\columnwidth]{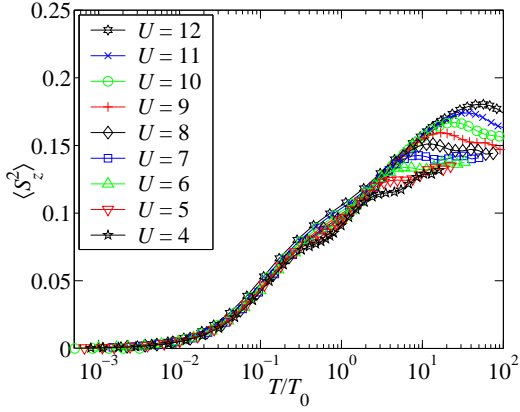}

   \caption{(Color online) The temperature dependence of the effective moment
     $\expect{S^2_z}_\textrm{imp}(T)$ versus $T/T_0$ for various
     values of $U$ and a fixed filling of  $n_\textrm{tot}=1.6$.
     Parameters: as in Fig.~\ref{fig:energy-scales}.
     }

   \label{fig:sz2-vs-x}
\end{figure}

Using Wilson's definition\cite{Wilson75} of an effective local moment,
$\expect{S^2_z}_\textrm{imp}(T)$ 
is given by the difference of  $\expect{S^2_z}$ of the effective SIAM
and the medium electron gas without impurity. Since the medium changes
with temperature, $\expect{S^2_z}_\textrm{imp}(T)$ is calculated for
each converged medium for the given temperature and $T^*$ is obtained by the
condition  $\expect{S^2_z}_\textrm{imp}(T^*)=0.05$. 

The results for all four different scales are depicted  in Fig.\
\ref{fig:energy-scales} using a fixed filling $n_\textrm{tot}=1.6$.
While the fit to the $T^2$ behavior contains a significant error bar due to
the logarithmic temperature scale of the NRG and the limited
temperature range of the Fermi-liquid regime, all other energy scales
can be obtained with high accuracy due to their definition. The
dimensionless coupling constant
\begin{equation}
  g=\rho_0 J = \frac{2 \Gamma_0 U}{\pi\abs{\e_f}\abs{\e_f+U}}
  \label{eq:coupling}
\end{equation}
has been estimated by the  Schrieffer-Wolff
transformation.\cite{SchriefferWol66}
All four scales show the same exponential dependency on $U$ with
$T_0 \propto\exp(-1/\alpha g)$ with $\alpha\approx0.8$ for large $U$
where the charge and low-temperature scales are well separated. This
indicates that, up to a filling-dependent proportionality constant,
there exists only \emph{one} low-temperature scale. Our findings are
in perfect agreement with the previously reported\cite{PruschkeBullaJarrell2000}
$T=0$ estimates of the low-temperature scale but in contrast to the
Gutzwiller results\cite{RiceUeda1986}  predicting an enhancement of the
$T_0/T_K= \exp(1/2g)$. However, the latter ratio is strongly filling
dependent as pointed out previously.\cite{PruschkeBullaJarrell2000}

As a consequence of this scaling analysis,
$\expect{S^2_z}_\textrm{imp}(T)$ is collapsed onto one master curve  
for a fixed filling $n_\textrm{tot}$ and different values of $U$ using the
scaling variable $x= T/T_0$ as displayed in
Fig.~\ref{fig:sz2-vs-x}. This master curve for
$\expect{S^2_z}_\textrm{imp}(T)$ shows slight deviation from its
corresponding shape for the SIAM with constant density of states;
\cite{KrishWilWilson80} however, it holds up to very high
temperatures. Clearly visible is the $U$ dependence of the maximum of
$\expect{S^2_z}_\textrm{imp}(T)$. Only in the local moment fixed point
the value of $\expect{S^2_z}_\textrm{imp} = 0.25$ is
reached.\cite{KrishWilWilson80} At high temperatures, deviations from
scaling are related to the crossover from the free orbital to the
local-moment fixed point\cite{KrishWilWilson80} and are independent of lattice
renormalization effects.

\begin{figure}
  \includegraphics[width=0.8\columnwidth]{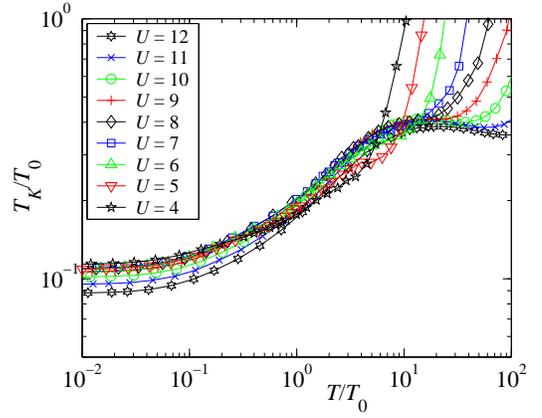}
  \caption{(Color online) The temperature-dependent Kondo temperature $T_K(T)$ of the
    effective site for various $U$ plots versus the scaling variable
    $x=T/T_0$ for fixed filling.
    Parameters: as in Fig.~\ref{fig:sz2-vs-x}.
    }
  \label{fig:TK-vs-x}
\end{figure}

In addition, we can define a ``temperature'' dependent Kondo scale $T_K(T)$ of
the DMFT effective site, by taking the DMFT medium density of states
$\rho_\textrm{eff}(\w)$ for a fixed temperature and solving the condition
$\expect{S^2_z}_\textrm{imp}(T_K)=0.05$ for such a fixed medium. Since
we would find a constant $T_K(T)$ for a SIAM, deviations from such a
constant function indicate lattice renormalization effects. This also
could give strong hints towards the existence of a second low-energy
scale. $T_K(T\to\infty)$ corresponds to a ``free-impurity'' Kondo
temperature. In the  Fermi-liquid regime, $T_K(T)$ approaches a
constant $T_K(0)$ measuring the lattice low-temperature scale:
$T_K(0)\sim T_0$.  In Fig.~\ref{fig:TK-vs-x}, the resulting
renormalized dimensionless Kondo temperature $T_K(T)/T_0$ is plotted
versus $T/T_K$ for different values of $U$. Even though two
distinguishable plateaus  are visible, indicating two different
low-temperature scales for large values of $U$, they are connected by
a fixed ratio independent of $U$. 

The existence of such universal scaling curves for
$\expect{S^2_z}_\textrm{imp}$ and $T_K(T)$ connecting these different
temperature regimes provides strong evidence that (i) the low-temperature
phase of the paramagnetic PAM is characterized by only one thermodynamic
low-temperature scale, and (ii) all additionally defined low-energy
scales assigned to features of some physical properties are
proportional to this single low-temperature scale. Our findings are in
agreement with the previous DMFT-NRG study\cite{PruschkeBullaJarrell2000}
as well as the series of DMFT-LMA investigations of scaling and transport
properties.\cite{LoganVid2003,VidLogan2004,LoganVid2005}
We did not find any indication for
a novel fixed point of the effective site introduced by the
nonconstant density of states. In the absence of a phase
transition, the effective site is characterized
by the same fixed points as the SIAM\cite{KrishWilWilson80} and,
therefore, only one universal low-temperature scale is expected.

\section{Conclusion and Outlook}
\label{sec:conclusion}

We presented a comprehensive study of the  transport properties of the
periodic Anderson model. We calculated the resistivity and the
thermoelectric power for metallic heavy-fermion systems as well as
Kondo insulators using  finite temperature DMFT-NRG and DMFT-MPT
approaches. While the DMFT-MPT gives extremely reliable results in the
weak correlation limit $\rho(0)U \ll 1$, it fails to reproduce the
correct low-energy scale for large values of $U$. In contrast, the
DMFT-NRG resolves accurately low-energy scales
for arbitrary parameters  but  provides limited spectral resolution
in finite-temperature spectral functions. The latter has made it very
challenging to obtain reliable transport properties since only
excitations of the order $T$ enter the transport integrals.
Nevertheless, we were able to use DMFT-NRG for the investigation of
transport properties as well as the four
different low-energy scales. All these energy scales are related by a
fixed but filling-dependent ratio, indicating the existence of only one
relevant low-temperature scale.

We conclude that all characteristic low-temperature transport properties of
heavy-fermion systems can be explained  within our numerical DMFT-NRG
treatment of the PAM. In the metallic situation one obtains a low residual
resistivity $\rho(0)$, a $\rho(T)$ rapidly increasing with increasing
temperature $T$ following a $T^2$ behavior for very low $T < T_0$, a maximum
of the order $100\;\mu\Omega\;\mathrm{cm}$ at a temperature $T_\textrm{max}$
(about $10$--$200\;\mathrm{K}$) and a $\rho(T)$ logarithmically
decreasing with increasing $T$ for $T > T_\textrm{max}$, thus showing
the typical experimental behavior.\cite{Scoboria79, Ott.75.84, Onuki87}
At the same time the thermoelectric power can exhibit several extrema,
for some parameters a change of sign, and large absolute values of the
magnitude of $50$--$150\;\mu\mathrm{V/K}$, as seen in
experiments.\cite{Steglich96,AndersHuth2001}
The optical conductivity shows a Drude peak and an additional midinfrared
peak for low $T$, whereas these structures merge to a broad structure at
high $T$. For Kondo insulators a crossover from the Kondo behavior at
$T > T_\textrm{max}$ to an activation behavior is obtained for $\rho(T)$
with a saturation due to the (artificial) finite imaginary part $\delta$
(simulating impurity scattering), and the optical conductivity has no Drude
peak for very low $T$. The absolute value of the thermoelectric power $S(T)$
can be even larger in or close to the Kondo insulator situation reaching giant
values of more than $200\;\mu\mathrm{V/K}$. The figure of merit can be
larger than 1, showing that these systems are candidates for
thermoelectric applications.

Thus far all our calculations did not include orbital degeneracy which
is subject to crystal electric field splittings. These additional
degrees of freedom are essential for the understanding of the
transport properties of heavy-fermion systems in the intermediate temperature
regime\cite{AndersHuth2001} as well as the broad midinfrared peak
observed in the optical conductivity.\cite{Degiorgi99} As recently
pointed out,\cite{KogaCox99,KogaZarandCox99,AndersPruschke2006} these
additional $f$ configurations can indeed introduce additional
low-energy scales due to competition or crossover between additional fixed
points. Apparently, the thermoelectric power is very sensitive to such
additional configurations since it is determined by particle-hole asymmetry
around the chemical potential. 

Therefore, in the future this work can and should be extended into different
directions. On the one hand, the inclusion of realistic $f$-shell
degeneracy and of crystal fields is highly desirable to come to a more
realistic description of the transport properties of heavy-fermion systems
valid also for high $T$, where the higher crystal field levels can be
excitated. On the other hand, the influence of disorder and alloying, in
particular of the substitution of the (magnetic) lanthanide or actinide ions
by nonmagnetic ions should be investigated and understood theoretically, as
there exist many experimental investigations systematically studying these
substitution effects.

\emph{Note added in proof.}
After completion of this work, potentially more accurate ways
of calculating the NRG spectral functions were
proposed\cite{WeichselbaumDelft2006,PetersPruschkeAnders2006} which might
increase the accuracy of future DMFT-NRG calculations.

\begin{acknowledgments}
We thank A.~Hewson, J.~Freericks, D.~Logan, R.~Bulla, N.~Grewe, and V.~Zlati\'c
for numerous discussions. C.G., F.B.A., and G.C.\ acknowledge financial support
by the Deutsche Forschungsgemeinschaft, Project No.\ \mbox{AN 275/5-1} and
funding of the NIC, Forschungszentrum J\"ulich, under Project No.\ HHB000.
\end{acknowledgments}


\end{document}